\documentclass[usenatbib,useAMS,twocolumn,fleqn]{mnras}
\usepackage[T1]{fontenc}
\usepackage{ae,aecompl}

\usepackage{amsmath,amstext,amsopn,amssymb}
\usepackage{txfonts}
\usepackage{bm}
\usepackage[artemisia]{textgreek}
\usepackage{graphicx}
\usepackage{float}

\newcommand{\secref}[1]{Section~\ref{#1}}
\newcommand{\figref}[1]{Fig.~\ref{#1}}
\newcommand{\eref}[1]{equation~(\ref{#1})}

\newcommand{\pcond}[2]{P({#1} \, \vert \, {#2})}
\newcommand{\vv}[1]{\bm{#1}}	
\newcommand{\vvd}[1]{\vv{#1}_{\text{d}}}	
\newcommand{\vvr}[1]{\vv{#1}_{\text{r}}}	
\newcommand{\vvdc}[1]{\vv{#1}_{\text{dc}}}	
\newcommand{\vvt}[1]{\vv{#1}_{\text{t}}}	
\newcommand{\mtx}[1]{\mathbfss{#1}}	
\newcommand{\mstellar}{M_{\text{stellar}}}  
\newcommand{\mbstar}{{m}_{B}^{\star}}  
\newcommand{\vmbstar}{\vv{m}_{B}^{\star}}  
\newcommand{\jab}{\mtx{J}_{(\alpha, \beta)}} 
\newcommand{\stmean}[1]{\langle{#1}\rangle} 
\newcommand{\logg}{\log_{10}}	
\DeclareMathOperator{\sk}{S}	
\DeclareMathOperator{\tr}{tr}
\DeclareMathOperator{\ud}{d}

\defcitealias{2014A&A...568A..22B}{B14}

\title[Bayesian graphs for SN Ia analysis]{Application of Bayesian graphs
to SN Ia data analysis and compression}

\author[C.~Ma, P.-S.~Corasaniti \& B.~A.~Bassett]
{Cong Ma,$^{1, 2,
3}$\thanks{Email: \href{mailto:cma@pmo.ac.cn}{\texttt{cma@pmo.ac.cn}}}
Pier-Stefano Corasaniti,$^3$ and Bruce A.~Bassett$^{4, 5, 6}$\\
$^1$Purple Mountain Observatory, Chinese Academy of Sciences, 2 West
Beijing Rd, 210008 Nanjing, China\\
$^2$Graduate School, University of the Chinese Academy of Sciences, 19A Yuquan
Rd, 100049 Beijing, China \\
$^3$LUTH, UMR 8102 CNRS, Observatoire de Paris, PSL Research University,
Universit{\'e} Paris Diderot, 5 Place Jules Janssen, F-92190 Meudon, France\\
$^4$Department of Mathematics and Applied Mathematics, University of Cape Town,
Cross Campus Rd, Rondebosch 7700, South Africa\\
$^5$African Institute for Mathematical Sciences, 6--8 Melrose Rd, Muizenberg
7945, South Africa\\
$^6$South African Astronomical Observatory, Observatory Rd, Observatory 7925,
South Africa
}

\date{Accepted XXX. Received YYY; in original form ZZZ}
\pubyear{2016}

\begin{document}
\label{firstpage}
\pagerange{\pageref{firstpage}--\pageref{lastpage}}
\maketitle

\begin{abstract}
    Bayesian graphical models are an efficient tool for modelling complex data
    and derive self-consistent expressions of the posterior distribution of
    model parameters. We apply Bayesian graphs to perform statistical analyses
    of Type Ia supernova  (SN Ia) luminosity distance measurements from the
    joint light-curve analysis (JLA) data set. In contrast to the $\chi^2$
    approach used in previous studies, the Bayesian inference allows us to
    fully account for the standard-candle parameter dependence of the data
    covariance matrix.  Comparing with $\chi^2$ analysis results, we find a
    systematic offset of the marginal model parameter bounds.  We demonstrate
    that the bias is statistically significant in the case of the SN Ia
    standardization parameters with a maximal $6\sigma$ shift of the SN
    light-curve colour correction. In addition, we find that the evidence for
    a host galaxy correction is now only $2.4\sigma$.  Systematic offsets on
    the cosmological parameters remain small, but may increase by combining
    constraints from complementary cosmological probes. The bias of the
    $\chi^2$ analysis is due to neglecting the parameter-dependent
    log-determinant of the data covariance, which gives more statistical
    weight to larger values of the standardization parameters.  We find a
    similar effect on compressed distance modulus data.  To this end, we
    implement a fully consistent compression method of the JLA data set that
    uses a Gaussian approximation of the posterior distribution for fast
    generation of compressed data.  Overall, the results of our analysis
    emphasize the need for a fully consistent Bayesian statistical approach in
    the analysis of future large SN Ia data sets.
\end{abstract}

\begin{keywords}
    methods: data analysis -- methods: statistical -- supernovae: general --
    cosmological parameters -- distance scale.
\end{keywords}

\section{Introduction}
\label{sec:intro}

Over the past decade, observational programmes dedicated to Type Ia supernovae
(SN Ia) have significantly enlarged the original data set that lead to the
pioneering discovery of the cosmic acceleration \citep{1998AJ....116.1009R,
1999ApJ...517..565P}. To date these systematic searches have detected about a
thousand SN Ia across a large redshift range
\citep[see][]{2006A&A...447...31A, 2007ApJ...659...98R, 2007ApJ...666..694W,
    2008AJ....135..338F, 2009ApJ...700..331H, 2010AJ....139..519C,
2012ApJ...750...99T, 2012ApJ...746...85S, 2013ApJ...763...88C}.  Thanks to
this new generation of SN surveys, it has been possible to achieve
unprecedented high statistical precision on luminosity distance measurements.
In fact, there is a widespread consensus that current cosmological constraints
from SN Ia are limited by systematic uncertainties
\citep[see][]{2011ApJS..192....1C, 2014ApJ...795...45S}.  Potential sources of
bias arise from variations of SN magnitudes that correlate with host galaxy
properties \citep[see][]{2010ApJ...715..743K, 2010MNRAS.406..782S,
2012MNRAS.426.2359M} as well as model assumptions in the light-curve fitting
methods that are used to standardize the SN sample. 

Recently, in an effort to bring SN Ia observations from different data sets on
a common ground, \citet[][hereafter
\citetalias{2014A&A...568A..22B}]{2014A&A...568A..22B} have performed a joint
light-curve analysis (JLA) of data from the Supernova Legacy Surveys (SNLS),
the Sloan Digital Sky Survey-II supernova survey (SDSS-II) and a variety of
programmes that targeted low- and high-redshift SNe. The full data set has
been made publicly available, including light-curve fitting parameters with
their covariance matrices and a compressed set of distance modulus data, thus
providing all elements necessary to perform statistically robust cosmological
data analyses.

SN Ia magnitudes are standardized using an empirical relation between the
maximum absolute magnitude peak and the time width \citep{1993ApJ...413L.105P,
1996AJ....112.2391H, 1999AJ....118.1766P} of the light curve and the SN colour
\citep{1998A&A...331..815T}. These parameters are first extracted for each SN
by fitting the observed light curves, then they are used in the
standard-candle relation to estimate the distance moduli from which
cosmological parameter constraints are finally inferred. This is the
operational mode of the SALT2 light-curve fitting model
\citep{2014ApJ...793...16M} originally introduced in
\citet{2007A&A...466...11G} and used to derive the measurements of
\citetalias{2014A&A...568A..22B}. A critical aspect of this process concerns
the propagation of uncertainties in the standardization parameters that
parametrize light-curve features in the standard-candle relation. In the
context of Bayesian statistics this problem is addressed unambiguously by
assigning priors to the standardization parameters. More in general, Bayesian
methods can handle all the complexity of large SN data sets while providing a
self-consistent probabilistic modelling of the data. As an example, in
$\chi^2$ analyses the residual SN intrinsic magnitude scatter is usually
fitted together with the cosmological parameters under the (unphysical)
requirement that the $\chi^2$ per degree of freedom of the best-fitting model
is $\sim 1$. This is not needed in the Bayesian framework where it is possible
to derive the full posterior probability distribution of the intrinsic
scatter.  \citet{2011MNRAS.418.2308M,2014MNRAS.437.3298M} have shown this to
be the case using a Bayesian hierarchical (or graphical/network) model of the
SN data.  Recently, \citet{2016ApJ...827....1S} have also applied a similar
formalism to the analysis of the JLA data set to simultaneously infer
constraints on cosmological and standardization parameters. Bayesian graphs,
also known as Bayesian networks, have a twofold advantage over $\chi^2$
statistical methods \citep[see for a review][]{Jensen2001}. On the one hand,
it provides a better understanding of the data through a graphical
representation of the causal and probabilistic connections of all problem's
variables. On the other hand, the graphical model allows one to directly
derive the factorized form of joint probability distributions for the
parameter of interests, thus providing a (numerical) solution even when the
problem is extremely complex. 

Here, we use Bayesian graphical models for the JLA data set to perform a
self-consistent cosmological parameter inferences that account for the
light-curve fitting parameter dependence of the data covariance. This is an
important point that has been overlooked in previous SN studies. We will show
that such a dependence not only impacts the cosmological parameter constraints,
but also the estimation of the standard-candle parameters. Neglecting such a
dependence is even more problematic in the case of compressed SN data sets.
Due to the statistical nature of the compression method, the effect of the
parameter-dependent covariance can lead to biased results. Once the
compression is done, there is no simple method to amend the inconsistency
using compressed data alone. 

The paper is organized as follows. In \secref{sec:cos-snia-data}, we introduce
the basic concepts of SN cosmology and describe the JLA data. In
\secref{sec:graph-mod}, we introduce Bayesian graphical models and discuss
their application of the JLA data set, while in \secref{sec:jlacosmo} we will
present the result of the cosmological parameter inference.  In
\secref{sec:datacompress}, we describe the statistical compression method
applied to the JLA distance modulus data and discuss the result of various
tests in \secref{sec:jla-comp-anls}. Finally, \secref{sec:conc} presents our
conclusion.

\section{Cosmology with SN Ia data}
\label{sec:cos-snia-data}

In this section, we will briefly review the main concepts of SN cosmology and
introduce the JLA data set. We will do so from our Bayesian point of view
which refers to the fact that (1) all parameters are considered random
variables to which we assign prior probability distributions based on
available information or the lack thereof; (2) data are described by random
variables with only one realization deduced from the observations, thus
formally described in probabilistic expression as conditioned variables.

Let us begin with the definition of distance modulus and consider an
astrophysical source with absolute magnitude $M_{\text{abs}}$ and apparent
magnitude $m$. The luminosity distance $d_{\text{L}}$ to the source can be
obtained from the distance modulus:
\begin{equation}
    \label{eq:mod_dist}
    \mu \equiv 5 \logg \left( \frac{d_{\text{L}}}{10~\text{pc}} \right) =
    m - M_{\text{abs}}.
\end{equation}
In the case of SN Ia, the observed magnitude (as measured at the peak
luminosity of the light curve) varies from one object to another.
Nevertheless, using correlations with other measurable features in the SN
light curve, it is possible to deduce a standard value that has been shown to
have a very small scatter over a large SN sample.

The possibility of this standardization was first pointed out by
\citet{1993ApJ...413L.105P} who showed that the SN peak luminosity correlates
with the rate of brightness decline (or stretch) of the light-curve.
Subsequently, \citet{1998A&A...331..815T} showed an additional correlation with
the SN colour.  Further correlations have been found with the host galaxy
properties, such as the star formation rate and metallicity
\citep{2005ApJ...634..210G, 2015ApJ...802...20R}, stellar mass
\citep{2010ApJ...715..743K} and galaxy morphology \citep{2009ApJ...700..331H}.
SN samples such as the JLA data set include such corrections as we shall
describe next.

\subsection{Description of the JLA data set}
\label{sec:jladesc}

The JLA data set presented in \citetalias{2014A&A...568A..22B} consists of 740
SN Ia measurements of the peak apparent $B$-band magnitude $\mbstar$ in the AB
magnitude system, the `stretch' or shape parameter of the light curve ${X}_1$
and the colour parameter ${C}$. These data can be represented in a concise
form by introducing the joint data vector
\begin{equation}
    \vv{v} = \vmbstar \oplus \vv{X}_1 \oplus \vv{C} =
	 \begin{bmatrix}
	     \vmbstar \\
	     \vv{X}_1 \\
	     \vv{C}
	 \end{bmatrix}.
    \label{eq:v}
\end{equation}
This is a $(3 \times 740 = 2220)$-dimensional random vector, and its
`realized' value is given by the JLA data reduction process.  The vector
$\vv{v}$ here contains all the elements of the data vector denoted by
$\vv{\eta}$ in \citetalias{2014A&A...568A..22B}, but transformed by a
permutation so that the block structure in \eref{eq:v} is maintained. The data
also provide the covariance matrix $\mtx{C}_{\vv{v}}$, a $2200 \times 2200$
symmetric positive-definite matrix that can be represented as a $3 \times 3$
block matrix
\begin{equation}
    \mtx{C}_{\vv{v}} = 
    \begin{pmatrix}
	\mtx{C}_{mm} & \mtx{C}_{mX} & \mtx{C}_{mC} \\
	\ddots       & \mtx{C}_{XX} & \mtx{C}_{XC} \\
	\ddots       & \ddots       & \mtx{C}_{CC}
    \end{pmatrix},
    \label{eq:cv}
\end{equation}
where each block is a $740 \times 740$ square matrix.  The matrix
$\mtx{C}_{\vv{v}}$ corresponds to the matrix $\mtx{C}_{\vv{\eta}}$ in
equation~11 of \citetalias{2014A&A...568A..22B} but with two slight
differences.  First, the matrix $\mtx{C}_{\vv{\eta}}$ is also permuted in the
columns and rows so that it conforms to the block structure in \eref{eq:cv}.
Secondly, the three additional diagonal components in equation~13 of
\citetalias{2014A&A...568A..22B}, i.e.\ the peculiar velocity, weak lensing
and other intrinsic dispersions, are added to the block $\mtx{C}_{mm}$ without
altering the computation of the covariance of the distance modulus vector
$\vv{\mu}$. It is worth stressing that the covariance $\mtx{C}_{\vv{v}}$
provided by \citetalias{2014A&A...568A..22B} already includes an estimate of
the intrinsic SN magnitude dispersion inferred from a cosmological parameter
independent restricted likelihood analysis described in their section~5.5.
Because of this, we do not consider an additional $\sigma_{\text{int}}$
dispersion parameter term as done for instance in the analysis of
\citet{2012A&A...541A.110L}.

\begin{figure}
    \includegraphics[width=\columnwidth]{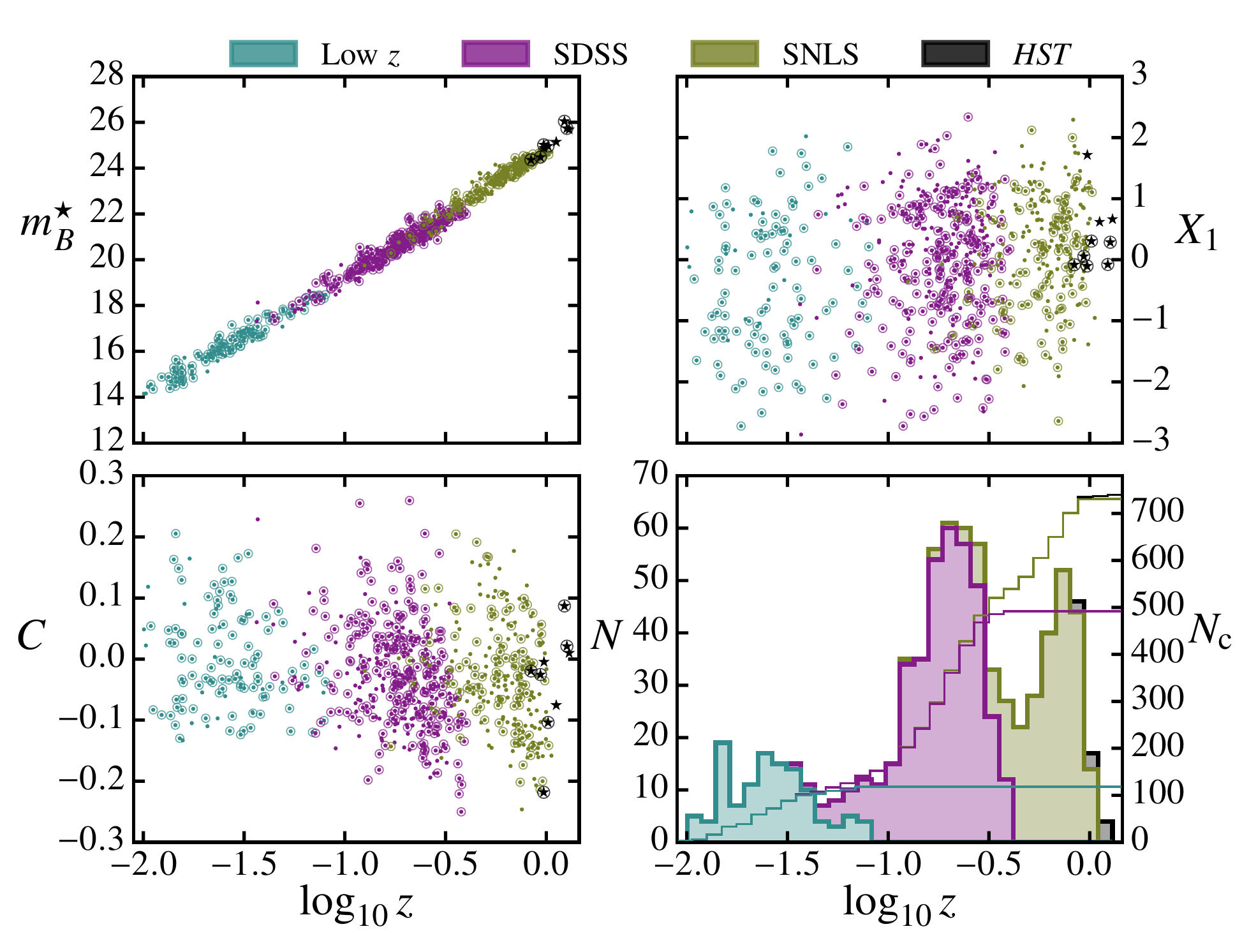}
    \caption{Visualization of the JLA data set coloured by the four component
	source surveys (Low $z$, SDSS, SNLS and \textit{HST}). Displayed are
	the apparent peak magnitude $\mbstar$ (top-left panel), the
	light-curve stretch $X_1$ (top-right panel) and the colour $C$
	(bottom-right panel). The bottom-right panel shows the SN redshift
	empirical distribution (left axis, filled histograms) and the
	cumulative distribution (right axis, unfilled histograms).}
    \label{fig:visualjla}
\end{figure}

These data are provided with a set of metadata containing non-random
information such as the SN redshift (for which errors are negligible), the
stellar mass of the host galaxy $\mstellar$ (in units of solar mass
$\mathrm{M}_{\sun}$) and a tag specifying the sample of origin of each SN. The
panels in \figref{fig:visualjla} summarize the redshift distribution of the
JLA observables.

Following the convention of \citetalias{2014A&A...568A..22B}, the SN sample is
standardized using the corrected apparent magnitude relation:
\begin{equation}
    \vvd{m} = \vmbstar + \alpha \vv{X}_1 + \beta \vv{C},
    \label{eq:tripp}
\end{equation}
which includes the stretch and colour corrections. Notice that such a
correction is made on the apparent magnitude \citep[see
also][]{2010MNRAS.406..782S} and not the absolute one as in the original paper
by \citet{1998A&A...331..815T}.  Nevertheless, this is merely a matter of
convention that has no effect on the data analysis.  In addition, the
parameter $\beta$ in \eref{eq:tripp} differs from the common usage
\citepalias[e.g.][]{2014A&A...568A..22B} by a minus sign with no effect on the
final results.

It is convenient to rewrite the linear combination \eref{eq:tripp} in terms of
a linear operator $\jab$ represented by an $n \times 3n$ rectangular block
matrix of three $n \times n$ blocks:
\begin{equation}
    \begin{split}
	\jab &=
	\begin{pmatrix}
	    \mtx{I} & \alpha \mtx{I} & \beta \mtx{I}
	\end{pmatrix}\\
	&=
	\begin{pmatrix}
	    1 &        &   & \alpha &   &   & \beta &    & \\
	      & \ddots &   &   & \ddots &   &   & \ddots & \\
	      &        & 1 &   &   & \alpha &   &   & \beta 
	\end{pmatrix},
    \end{split}
\end{equation}
where $\mtx{I}$ is the $n \times n$ identity matrix and $n$ is the number of
data points (740). Thus, \eref{eq:tripp} can be written as
\begin{equation}
    \vvd{m} = \jab \vv{v}.
\end{equation}

The rest of the standardization accounts for galaxy host-dependent corrections.
In \citetalias{2014A&A...568A..22B}, the authors adopted a one-step correction
to the absolute magnitude of each SN as given by
\citep[seealso][]{2012ApJ...746...85S}
\begin{equation}
    M_{\text{d}} ( \Delta_M ) =
    \begin{cases}
	M^{1}_B            & \text{if $\mstellar <
	10^{10}~\mathrm{M}_{\sun}$,} \\
	M^{1}_B + \Delta_M & \text{otherwise,}
    \end{cases}
    \label{eq:absmagcorr}
\end{equation}
where $M^{1}_B$ and $\Delta_M$ are global random variables fitted with the
cosmological parameters. Notice that the SN absolute magnitude $M^{1}_B$ acts
as a free offset in the distance modulus. However, due to the degeneracy with
the Hubble constant entering in the luminosity distance, as seen in
\eref{eq:thlumindist}, $M^{1}_B$ is usually unconstrained unless additional
external information is included in the analysis (a point to which we will
come back at the end of \secref{sec:lumpre}). Alternatively, one can fix
$M^{1}_B$ to an arbitrary value and compensate the loss of randomness by
introducing another suitable random variable in the analysis. This is the
solution that we adopt here and set its value $M^{1}_B = -19.05$ as quoted in
\citetalias[appendix E]{2014A&A...568A..22B}.

Notice that the standardization parameters $\alpha$, $\beta$, and $\Delta_M$
are again random variables.  In general, their joint distribution is assumed
to be the same for all individual SNe in the sample.  Being random variables,
in the Bayesian approach they can be assigned a prior probability based on
prior information (or the lack thereof), and their posterior probability can
be obtained by the inference process.  Therefore, in this paper, we will not
attempt to fix them at any particular values.

To summarize, the distance modulus data vector reads as
\begin{equation}
    \label{eq:mudata}
    \vvd{\mu} = \vvd{m} - \vvd{M} = \jab \vv{v}-\vvd{M}( \Delta_M ).
\end{equation}
This can be seen as the result of a parameter-dependent affine transformation
of the JLA-reduced data vector $\vv{v} = \vmbstar \oplus \vv{X}_1 \oplus
\vv{C}$.  Assuming $\vv{v}$ to be Gaussian-distributed, for given
standardization parameter vector values $\vv{\varphi} = (\alpha, \beta,
\Delta_M)$, $\vvd{\mu}$ is also Gaussian-distributed with covariance
\begin{equation}
    \mtx{C}_{\text{d}} = \jab \mtx{C}_{\vv{v}} \jab^{\intercal}
    \label{eq:mudatacov}
\end{equation}
conditional on $\vv{\varphi}$.  For completeness, we remark that even
if $\vv{v}$ is not a Gaussian-distributed variable, the covariance of
$\vvd{\mu}$ is still given by \eref{eq:mudatacov}, although in this case
higher moments than the second order are required to fully characterize the
distribution.

\subsection{Cosmological model of the luminosity distance}
\label{sec:lumpre}

In a Friedman--Lema{\^i}tre--Robertson--Walker (FLRW) background the
luminosity distance $d_\text{L}$ at redshift $z$ reads as \citep[see
e.g.][]{1999astro.ph..5116H}
\begin{equation}
    d_{\text{L}}(z) = \frac{c}{H_0} (1 + z) \sk_k \left[ \int_0^z \frac{\ud
    z'}{E(z')} \right],
    \label{eq:thlumindist}
\end{equation}
where $H_0$ is the Hubble constant, $c$ is the speed of light and the function
$\sk_k$ depends on the curvature parameter $\Omega_k$,
\begin{equation}
    \sk_k \left( \cdot \right)=
	\begin{cases}
	    \frac{1}{\sqrt{\Omega_k}} \sinh \left[ \sqrt{\Omega_k} \left( \cdot
	    \right) \right] & \Omega_k > 0, \\
	    \cdot \quad \text{i.e.\ identity} & \Omega_k = 0, \\
	    \frac{1}{\sqrt{-\Omega_k}} \sin \left[ \sqrt{-\Omega_k} \left(
	    \cdot \right) \right] & \Omega_k < 0.
	\end{cases}
\end{equation}
The function $E(z)$ is the dimensionless expansion rate given by
\begin{equation}
    E(z) = \left[ \Omega_{\text{M}} (1 + z)^3 + \Omega_k (1 + z)^2 +
    \Omega_{\text{DE}}f_{\text{DE}}(z) \right]^{\frac{1}{2}},
    \label{eq:esolution}
\end{equation}
where $\Omega_{\text{M}}$ and $\Omega_{\text{DE}}$ are the matter and dark
energy density, respectively (with $\Omega_k = 1 - \Omega_{\text{M}} -
\Omega_{\text{DE}}$), and $f_{\text{DE}}(z)$ is a function characterizing the
dark energy density evolution. For a dark energy component with
redshift-dependent equation of state $w(z)$ this reads as
\begin{equation}
    f_{\text{DE}}(z) = \exp \left\{ 3 \int_0^z \left[ 1 + w(z') \right] \ud
    \ln \left( 1 + z' \right) \right\}.
\end{equation}

Here, it is worth noticing that the exact numerical value of the cosmological
distance at a given redshift and for a given set of cosmological parameters
depends on the choice of physical units. In contrast, the distance modulus
$\vvd{\mu}$ given by \eref{eq:mudata} depends on the absolute magnitude
$M^{1}_B$ previously discussed. Hence, equations~(\ref{eq:mod_dist}) and
(\ref{eq:mudata}) differ by an unknown magnitude-calibration constant $M$:
\begin{equation}
    M \equiv \mu_{\text{d}} - \mu.
\end{equation}
In the Bayesian approach, this is treated as a global random variable to be
fitted against the data. In such a case, $M$  accounts for any deviation of
the predicted value of the distance modulus from the observed one (due to the
specific choice of the value of $M^{1}_B$).

As already mentioned, any constant offset in the standard-candle relation is
degenerate with the value of $H_0$ entering the luminosity distance; thus,
$H_0$ and $M$ can be re-absorbed into a single parameter
\begin{equation}
    M' = M - 5 \logg h,
\end{equation}
where $h = H_0 / (100~\text{km s}^{-1}~\text{Mpc}^{-1})$ is the dimensionless
Hubble constant.  However, from our Bayesian perspective, $M$ and $H_0$ are
indeed different, since they may have different prior probabilities. In
particular, the prior distribution for $H_0$ can be inferred from observations
\citep[e.g.][]{2014MNRAS.440.1138E, 2015arXiv150201589P}, usually a fairly
localized Gaussian distribution. On the other hand, having no prior
information on $M$, we assigned a uniform prior.  We refer the readers to
\secref{sec:jlacosmo} for a detailed discussion of the degeneracy of $M$ and
$H_0$ and the joint constraints from the SN data analysis.

\section{Graphical models}
\label{sec:graph-mod}

Here, we introduce Bayesian graphical models (or networks) and describe their
application to SN Ia data. The literature on graphical representations of
Bayesian statistical models is quite vast; we refer the interested reader to
review papers by \citet{graphical} and \citet{2005physics..11182D} for a
first introduction and to \citet{km} for an extended treatment of the subject.

\subsection{Statistical inference and graphical representations}

We illustrate the use of Bayesian graphs with a simple toy model. Let us
consider a distance modulus data vector $\vv{\mu}$ with covariance matrix
$\mtx{C}$ and a theoretical model specified by the parameter vector
$\vv{\varTheta}$ predicting the distance modulus through a deterministic
function of the model parameters, i.e.\ $\vvt{\mu} =
f_{\text{t}}(\vv{\varTheta})$, such as the FLRW cosmic expansion model of
\eref{eq:thlumindist}.  We want to infer the posterior probability density
function (PDF) of the model parameters given the observations,
$\pcond{\vv{\varTheta}}{\vv{\mu}, \mtx{C}}$.  Using the definition of marginal
probability this reads as:
\begin{equation}
    \pcond{\vv{\varTheta}}{\vv{\mu}, \mtx{C}} =
    \int \pcond{\vv{\varTheta}, \vvt{\mu}}{\vv{\mu}, \mtx{C}} \ud \vvt{\mu},
    \label{eq:marginal}
\end{equation}
where the integrand is given by the definition of conditional probability
density
\begin{equation}
    \pcond{\vv{\varTheta}, \vvt{\mu}}{\vv{\mu}, \mtx{C}} =
    \frac{P(\vv{\varTheta}, \vvt{\mu}, \vv{\mu}, \mtx{C})}{P(\vv{\mu},
    \mtx{C})}.
    \label{eq:bayesrule}
\end{equation}
The term in the numerator is the joint probability distribution which can be
expressed as a factorization of conditional probabilities using the chain rule:
\begin{equation}
    P(\vv{\varTheta}, \vvt{\mu}, \vv{\mu}, \mtx{C}) =
    \pcond{\vv{\mu}}{\vvt{\mu}, \mtx{C}} P(\mtx{C}) \pcond{\vvt{\mu}}{
    \vv{\varTheta}} P(\vv{\varTheta}).
    \label{eq:chain}
\end{equation}

\begin{figure}
    \centering
    \includegraphics[width=.44\columnwidth]{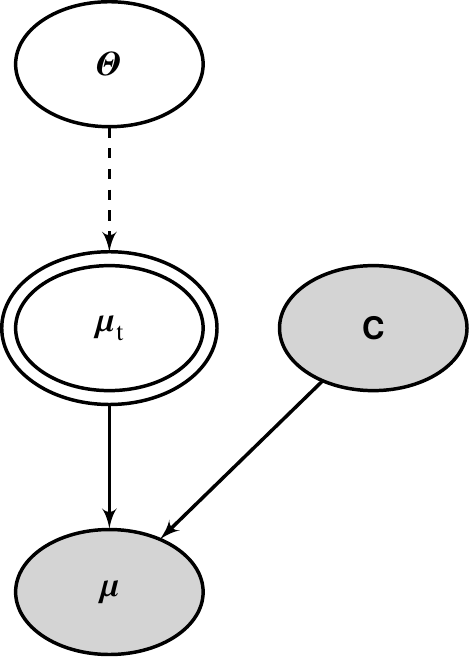}
    \caption{Graphical model representing deterministic and probabilistic
	relations between model parameters ($\vv{\varTheta}$), model
	predictions ($\vvt{\mu}$) and variables with evidence ($\vv{\mu},
	\mtx{C}$), in this case deduced from observational data.}
    \label{fig:bgproto}
\end{figure}

The dependence relations between all the variables of the problem as expressed
in the above equation can be represented in the graphical model shown in
\figref{fig:bgproto}. This is a directed acyclic graph (DAG) in which each
variable is represented by a node, while its relation to other variables is
marked by edges connecting the corresponding nodes. Deterministic relations
are represented by dashed edges, while solid edges indicate probabilistic
relations.

In \figref{fig:bgproto} we can already identify different kinds of nodes.
First, the grey nodes represent the random variables on which we have
evidence.  The evidence may come in the form of observational data or other
considerations specified probabilistically.  Henceforward we will denote these
nodes as `evident' ones, which is more general than the term `observed', thus
avoiding confusion with purely observed data.  Secondly, a node may be marked
by a double-circled boundary if it is deterministic, i.e.\ its conditional
distribution is a Dirac $\delta$ distribution.  In this paper, our notations
for the nodes follow that of \citet{2013arXiv13017412S}.

Notice that both the nodes for $\mtx{C}$ and $\vv{\varTheta}$ have
no parents, i.e.\ there are no edges from other nodes leading to them.  In
this sense, both may be said to be `unconditioned'.  However, they take
different roles in the statistical reasoning.  The theoretical model parameter
$\vv{\varTheta}$ is directly specified by its prior probability
$P(\vv{\varTheta})$; on the other hand, $\mtx{C}$ is an evident variable.  In
fact, though it may not be a direct observable, it can be derived by a
data-processing pipeline that propagates the statistical distributions from a
variety of observables.  We can thus imagine another graphical model in which
edges that flow from the observables ultimately arrive at $\mtx{C}$.  However,
once this upstream analysis is performed and $\mtx{C}$ is given its evident
value, its use in a subsequent analysis severs the links to the original
observables in the `upstream' of the pipeline \citep[][chapter 2.5.1]{km}.
Thus, while $\mtx{C}$ has evidence, the parameter $\vv{\varTheta}$ has none
which justify our convention in denoting their respective nodes.  The data
variable $\vv{\mu}$ occupies the root node of the graph. If there are more
data sets available, these will appear as multiple root nodes at the bottom of
the graph.

Starting from the graphical model in \figref{fig:bgproto}, one can easily
construct the factorized joint probability distribution by traversing the
graph, which is equivalent to the chain rule. Each non-observed starting node
contributes with a prior PDF [e.g.\ $\vv{\varTheta} \to P(\vv{\varTheta})$],
while non-starting nodes contributes with conditional probabilities that are
conditional to the variables associated with the connected nodes [e.g.\
$\vv{\mu} \to \pcond{\vv{\mu}}{\vvt{\mu}, \mtx{C}}$].  The grey, `observed'
nodes provide evidence, usually in the form of the available realization
provided by the data set, which constrains the randomness of the parameters.

Before evaluating the posterior distribution $\pcond{\vv{\varTheta}}{\vvt{\mu},
\mtx{C}}$, we first need to specify the form of the various terms in
\eref{eq:chain}.  In the case of Gaussian-distributed data, we have 
\begin{equation}
    \pcond{\vv{\mu}}{\vvt{\mu}, \mtx{C}} =
    \frac{\exp \left[ -\frac{1}{2} \left( \vv{\mu} - \vvt{\mu}
	    \right)^{\intercal} \mtx{C}^{-1} \left( \vv{\mu} - \vvt{\mu} \right)
    \right]}{\sqrt{(2 \upi)^n \det \mtx{C}}},
\end{equation}
where $n$ is the dimension of the vector random variable or the number of data
points. Since there is no uncertainty in the theoretical model prediction of
distance modulus, the conditional probability is a $\delta$ distribution
\begin{equation}
    \pcond{\vvt{\mu}}{\vv{\varTheta}} =
    \delta[ \vvt{\mu} - f_{\text{t}}(\vv{\varTheta}) ].
\end{equation}
Substituting these expressions in \eref{eq:chain} and computing the integral in
\eref{eq:marginal}, we obtain the familiar expression of the posterior
distribution
\begin{equation}
    \pcond{\vv{\varTheta}}{\vv{\mu}, \mtx{C}} =
    \frac{1}{Z(\vv{\mu}, \mtx{C})} \mathcal{L}(\vv{\varTheta}; \vv{\mu},
    \mtx{C}) P(\vv{\varTheta}),
    \label{eq:posterior}
\end{equation}
where $Z(\vv{\mu}, \mtx{C})\equiv P(\vv{\mu}, \mtx{C}) / P(\mtx{C}) =
\pcond{\vv{\mu}}{\mtx{C}}$ is a normalization constant, usually dubbed as the
`Bayesian evidence' or `marginal likelihood' that is relevant for model
selection \citetext{see e.g.\ \citealt{1996ApJ...471...24J};
\citealt*{2004ApJ...617L...1B}; \citealt{2006MNRAS.369.1725M,
2007MNRAS.378...72T}} and
\begin{equation}
    \mathcal{L}(\vv{\varTheta}; \vv{\mu}, \mtx{C}) =
    \frac{\exp \left\{
	    -\frac{1}{2} \left[ \vv{\mu} - \vvt{\mu}(\vv{\varTheta})
	    \right]^{\intercal} \mtx{C}^{-1} \left[ \vv{\mu} -
	    \vvt{\mu}(\vv{\varTheta}) \right] \right\}}{\sqrt{(2 \upi)^n \det
    \mtx{C}}}
\end{equation}  
is the so-called Gaussian likelihood function. Taking the logarithm of
\eref{eq:posterior} we obtain
\begin{equation}
    \begin{split}
	\ln \pcond{\vv{\varTheta}}{\vv{\mu}, \mtx{C}} =
	&- \ln Z - \frac{n}{2} \ln (2 \upi) - \frac{1}{2} \ln \det \mtx{C} \\
	&- \frac{1}{2} \chi^2(\vv{\varTheta}) + \ln P(\vv{\varTheta}),
    \end{split}
\end{equation}
where
\begin{equation}
  \chi^2 =
  \left[ \vv{\mu} - \vvt{\mu}(\vv{\varTheta}) \right]^{\intercal}
  \mtx{C}^{-1}
  \left[ \vv{\mu} - \vvt{\mu}(\vv{\varTheta}) \right],
\end{equation} 
is the object of the $\chi^2$ analysis. The value of evidence variables in
\eref{eq:posterior}, in this case $\mtx{C}$ and $\vv{\mu}$, can then be fixed
at their realized values as given by the data set, and in this regard
\eref{eq:posterior} becomes a function of $\vv{\varTheta}$ that can be readily
evaluated or sampled.  For brevity, in this paper we will not make notational
distinction of conditioning variables and their realized values in the
equations.

\begin{figure}
    \centering
    \includegraphics[width=.44\columnwidth]{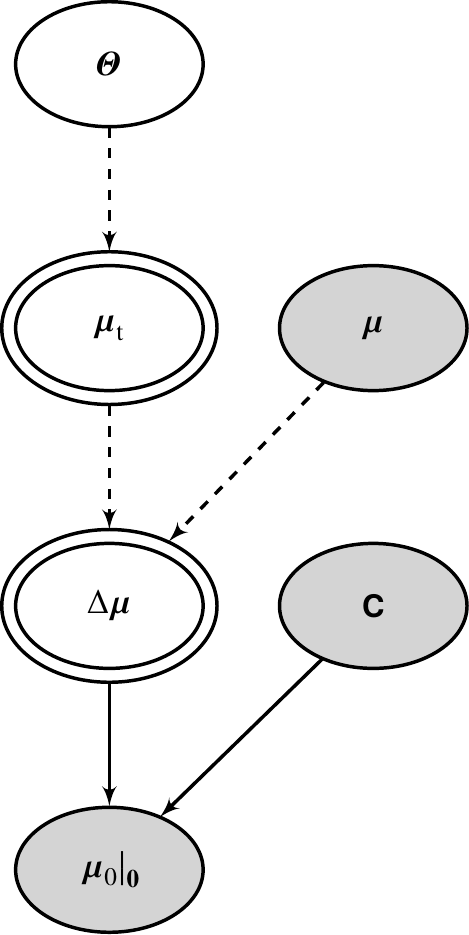}
    \caption{Graphical model as in \figref{fig:bgproto} with the addition of a
    deterministic node $\Delta \vv{\mu} = \vv{\mu} - \vvt{\mu}$ and an evident
    node $\vv{\mu}_0$ whose value is to be fixed at $\vv{0}$.}
    \label{fig:bgrotated}
\end{figure}

We would like to point out that the graphical model shown in
\figref{fig:bgproto} can be extended without altering the final posterior
calculation by adding a deterministic node for the variable $\Delta \vv{\mu} =
\vv{\mu} - \vvt{\mu}$ and an evident node $\vv{\mu}_0$, with the `realized
value' $\vv{0}$, as represented in \figref{fig:bgrotated}.  The
evidence on this node where the edges converge is not given by observational
data, but its specification is indispensable for the purpose of ensuring that
we are fitting the model $\vvt{\mu}$ to the data \citep[][chapter
2.5.3]{km}.  Then, the posterior distribution can be constructed similarly to
the earlier example as
\begin{equation}
    \begin{split}
	\pcond{\vv{\varTheta}}{\vv{\mu}_0, \vv{\mu}, \mtx{C}} =
	\frac{P(\vv{\varTheta})}{Z(\vv{\mu}_0,\ \vv{\mu}, \mtx{C})}
	\iint \left[ \pcond{\vv{\mu}_0}{\Delta \vv{\mu}, \mtx{C}}
	\pcond{\Delta \vv{\mu}}{\vv{\mu}, \vvt{\mu}} \right. \\
	\times \left. \pcond{\vvt{\mu}}{\vv{\varTheta}} \ud \vvt{\mu} \ud
	\Delta \vv{\mu} \right],
    \end{split}
    \label{eq:postmod}
\end{equation}
where $\pcond{\vv{\mu}_0}{\Delta\vv{\mu}, \mtx{C}}$ is given by a Gaussian
distribution with mean $\Delta \vv{\mu}$ and covariance $\mtx{C}$ and
$\pcond{\Delta\vv{\mu}}{\vv{\mu}, \vvt{\mu}} = \delta [ \Delta \vv{\mu} - (
\vv{\mu} - \vvt{\mu} ) ]$.  It is straightforward to see that performing the
integration in \eref{eq:postmod} yields the same result as
\eref{eq:posterior}, which assures us that the transformation does not alter
the results of statistical inference.

The extended graphical model may appear trivial.  However, the addition of the
extra nodes is indeed the key to handle the fact that the distance modulus
data from SN Ia depend on additional standardization parameters.  In
particular, both $\vv{\mu}$ and $\mtx{C}$ now occupy similar positions with no
parents, and our discussion about such data-derived evident variables,
exemplified by $\mtx{C}$ in the context of \figref{fig:bgproto}, now applies
symmetrically to both.

\subsection{Graphical model of inference with JLA data}

\begin{figure}
    \centering
    \includegraphics[width=.75\columnwidth]{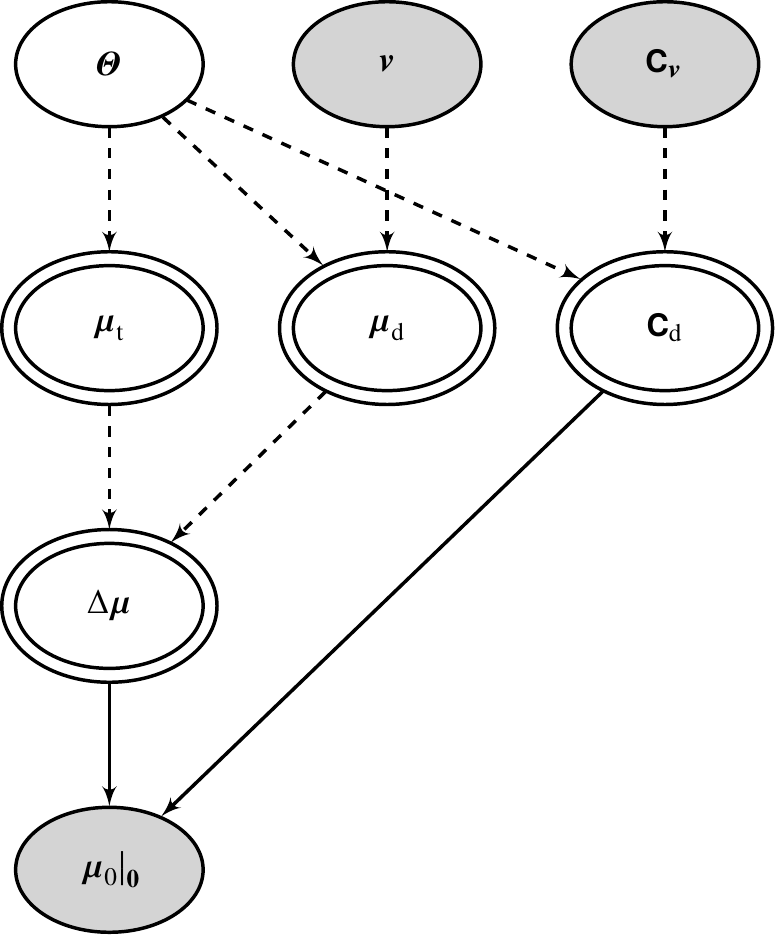}
    \caption{Graphical model of cosmological analysis using the JLA data set.}
    \label{fig:bgjla}
\end{figure}

In \secref{sec:jladesc} we have shown that the data vector
$\vv{\mu}_{\text{d}}$ and the covariance matrix $\mtx{C}_{\text{d}}$ are the
result of an affine transformation over the light-curve fitting parameters
data vector $\vv{v}$ and its covariance $\mtx{C}_{\vv{v}}$. The effect of this
affine transformation is to mix observed data and model parameters. In order
to disentangle them at the level of the calculation of the posterior
distribution, it is convenient, as in the case of the toy model shown in
\figref{fig:bgrotated}, to introduce the deterministic variable
$\Delta\vv{\mu} = \vv{\mu} - \vvt{\mu}$ and the evident one $\vv{\mu}_0 =
\vv{0}$. The Bayesian graphical model for the JLA data sets is shown in
\figref{fig:bgjla} where the parameter vector $\vv{\varTheta}$ now includes
the cosmological and the standard-candle relation parameters respectively.
From \figref{fig:bgjla} it is straightforward to derive the form of the joint
probability distribution and compute posterior distribution of the model
parameters given the data:
\begin{equation}
    \begin{split}
	\ln \pcond{\vv{\varTheta}}{\vv{\mu}_0,\ \vv{v}, \mtx{C}_{\vv{v}}} =
	&-\ln Z - \frac{n}{2} \ln (2 \upi) - \frac{1}{2} \ln \det
	\mtx{C}_{\text{d}}(\vv{\varTheta}) \\
	&- \frac{1}{2} \chi^2(\vv{\varTheta}) + \ln P(\vv{\varTheta}),
    \end{split}
    \label{eq:fullpost}
\end{equation}
where
\begin{equation}
    \chi^2(\vv{\varTheta}) =
    \left[ \vv{\mu}_{\text{d}}(\vv{\varTheta}) - \vvt{\mu}(\vv{\varTheta})
    \right]^{\intercal}
    \mtx{C}_{\text{d}}^{-1}(\vv{\varTheta})
    \left[ \vv{\mu}_{\text{d}}(\vv{\varTheta}) - \vvt{\mu}(\vv{\varTheta})
    \right]
    \label{eq:fullchi2}
\end{equation} 
with $\vv{\mu}_{\text{d}}(\vv{\varTheta})$ given by \eref{eq:mudata} and
$\mtx{C}_{\text{d}}$ from \eref{eq:mudatacov} as consequences of the affine
transformation which standardizes the SN Ia distance moduli.

The above expressions show an evident fallacy of the $\chi^2$ analysis,
namely neglecting the contribution of the parameter-dependent covariance.
This term cannot be dismissed as an implied constant even if one argues for the
use of $\chi^2$ statistics as motivated by the non-Bayesian theory of least
squares which yields the minimum variance unbiased estimator. In fact, this
theory requires that the covariance (or dispersion) matrix has to be known up
to a {\em constant} multiplier \citep{rao}. However, if the covariance is
parameter dependent, then in order to apply the classical least-squares
approach, the covariance must be approximated quadratically so that the
parameter-dependent contribution can be absorbed into a quadratic form with
constant dispersion matrix. Hence, such dependence does need to be properly
propagated in the final parameter inference. As we will show next, neglecting
this term can lead to biased results since maximizing the posterior is not
equivalent to minimizing the $\chi^2$.

\section{JLA cosmological parameter constraints}
\label{sec:jlacosmo} 

We perform a cosmological parameter inference using the JLA data set and
compare results based on the computation of the posterior distribution
\eref{eq:fullpost} versus the customary $\chi^2$-analysis specified by
\eref{eq:fullchi2}. We will refer to the former as the Bayesian approach and
the latter as `$\chi^2$'. As the target model we consider a flat dark energy
$w$CDM model with parameters $\vv{\varTheta} = [ \Omega_{\text{M}}, w, h, M,
\alpha, \beta, \Delta_M ]$, where $w$ is a constant dark energy equation of
state parameter.  We assume a Gaussian prior on $h$ with mean $0.688$ and
standard deviation $0.033$ consistent with the recent analysis on the
distances to nearby SN Ia by the Nearby Supernova Factory project
\citep{2015ApJ...802...20R}.  Following the discussion in
\citet[section~5.4]{2015arXiv150201589P}, we use the NFS value obtained from
an independent megamaser distance calibration to NGC~4258
\citep{2013ApJ...775...13H}. This result is consistent (within $0.5 \sigma$)
with the value of the Hubble constant obtained from the \textit{Hubble Space
Telescope} (\textit{HST}) Cepheid and nearby SN Ia data
\citep{2011ApJ...730..119R} as re-analysed by \citet{2014MNRAS.440.1138E} also
calibrated on NGC~4258 alone.  This prior differs from that used in the main
analysis of \citetalias{2014A&A...568A..22B} that assumed a hard prior $h =
0.7$ (while letting $M_B^1$ to freely vary).\footnote{As discussed at the end
of \secref{sec:lumpre}, fixing $h$ while letting $M_B^1$ to vary is not the
same as treating both parameters as random variables with different priors.
However, the cosmological parameter inference is insensitive to the choice of
a specific value of $h$ whether in the form of a hard prior or as a mean of a
Gaussian prior when the SN Ia data are used alone.} For the other parameters
we assume uniform priors in the following intervals: $\Omega_{\text{M}} \in
[0, 1]$, $w \in [-2.5, 1]$, $M \in [-5, 5]$, $\alpha \in[-1, 1]$, $\beta \in
[-10, 10]$ and $\Delta_M \in [-0.5, 0.5]$.

We evaluate the posterior distribution using the Markov chain Monte Carlo
(MCMC) method as implemented in by the
\textsc{pymc}\footnote{\url{https://pymc-devs.github.io/pymc/}} library
\citep*{JSSv035i04}. The $\chi^2$ analysis based on \eref{eq:fullchi2} is
performed by inserting a potential function proportional to $\sqrt{\det
\mtx{C}_{\text{d}}}$ in the joint probability distribution, which compensates
the term $(\ln \det \mtx{C}_{\text{d}}) / 2$ in \eref{eq:fullpost} so that the
`standard' $\chi^2$ analysis is emulated. We run four chains with $5\times
10^5$ samples each, and check their convergence using the Gelman--Rubin test
\citep{gelmanrubin, brooksgelman}. The estimated Monte Carlo standard error on
the parameter mean is of the order of $10^{-2}$ of statistical standard
deviation, negligibly affecting the results.

\begin{figure}
    \includegraphics[width=.9\columnwidth]{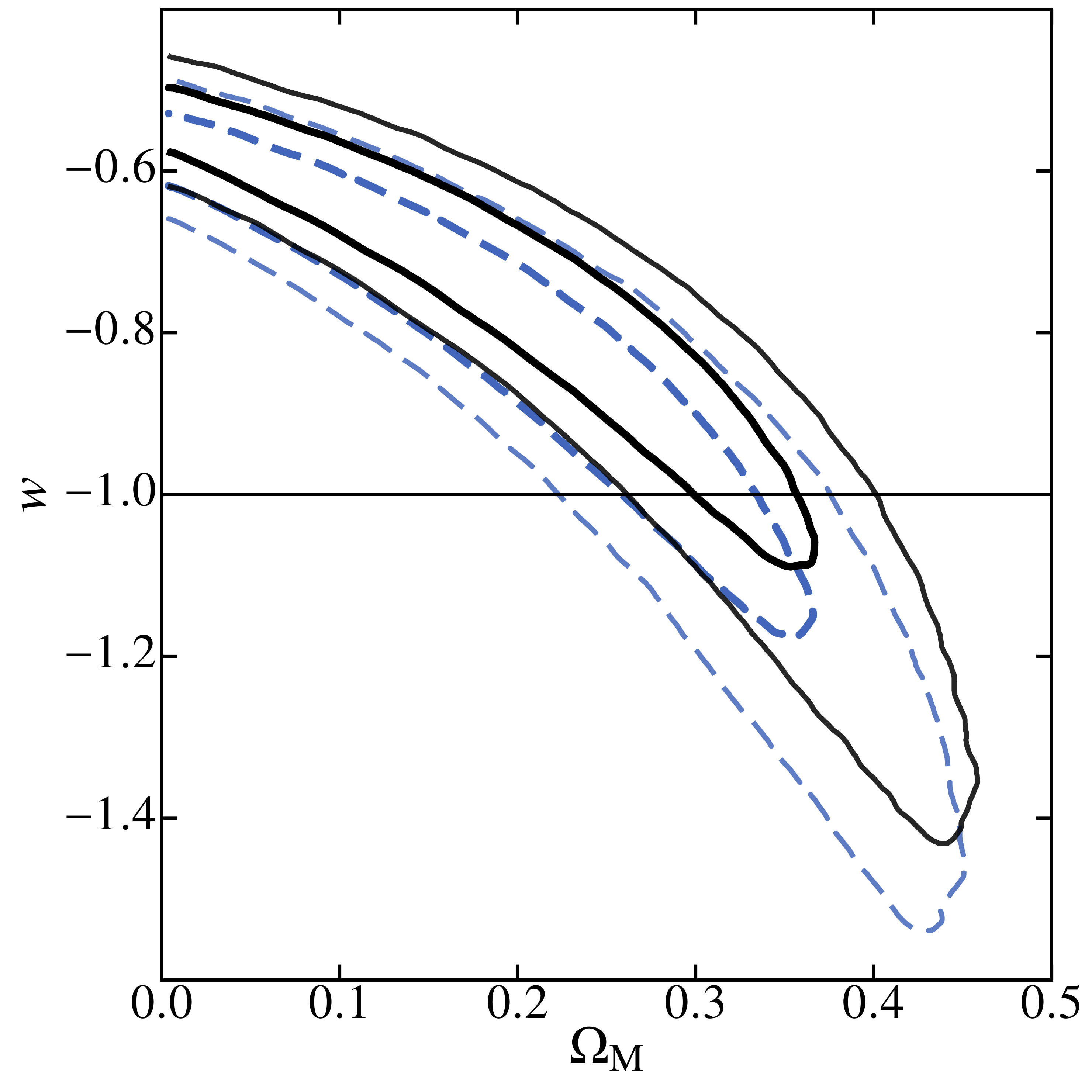}
    \caption{Marginal $0.683$ and $0.95$ two-dimensional credibility regions in
    the $\Omega_{\text{M}}$--$w$ plane for a flat $w$CDM model derived from
    the analysis of the full posterior distribution (black solid lines) and the
    $\chi^2$ analysis (blue dashed lines).}
    \label{fig:wcdm}
\end{figure}

In \figref{fig:wcdm} we plot the marginalized PDF contours in the
$\Omega_{\text{M}}$--$w$ plane obtained from the Bayesian and $\chi^2$
analyses, respectively. We can see that the effect of neglecting the
covariance term results in a systematic offset of the probability contours.
The marginalized mean and standard deviation from the MCMC samples give $w =
-0.82 \pm 0.22$ for the posterior PDF analysis and $w = -0.88 \pm 0.24$ for
the $\chi^2$ approach, while we find $\Omega_{\text{M}} = 0.22 \pm 0.11$ in
both cases. These results are consistent to within $1\sigma$ with the findings
of \citet{2016ApJ...827....1S}.

Although the bias effect on $w$ appears to be small (about $0.25 \sigma$) this
can be deceptive. As is well known, the parameters $(w, \Omega_{\text{M}})$
display significant degeneracy when using SN Ia data alone.  The cosmological
constraints noticeably tighten when combined with other cosmological probes as
shown for instance in \citetalias{2014A&A...568A..22B}.  Hence, the bias
effect shown here may be enhanced by the addition of complementary constraints
as also found by \citet{2016ApJ...827....1S}. The comparison of the marginal
distributions of $(w, \Omega_{\text{M}})$ is not all there is to the full
inference. Unsurprisingly, we find a more significant bias effect on $(\alpha,
\beta)$ and $\Delta_M$ on which $\ln \det \mtx{C}_{\text{d}}$ directly
depends.

\begin{figure}
    \includegraphics[width=\columnwidth]{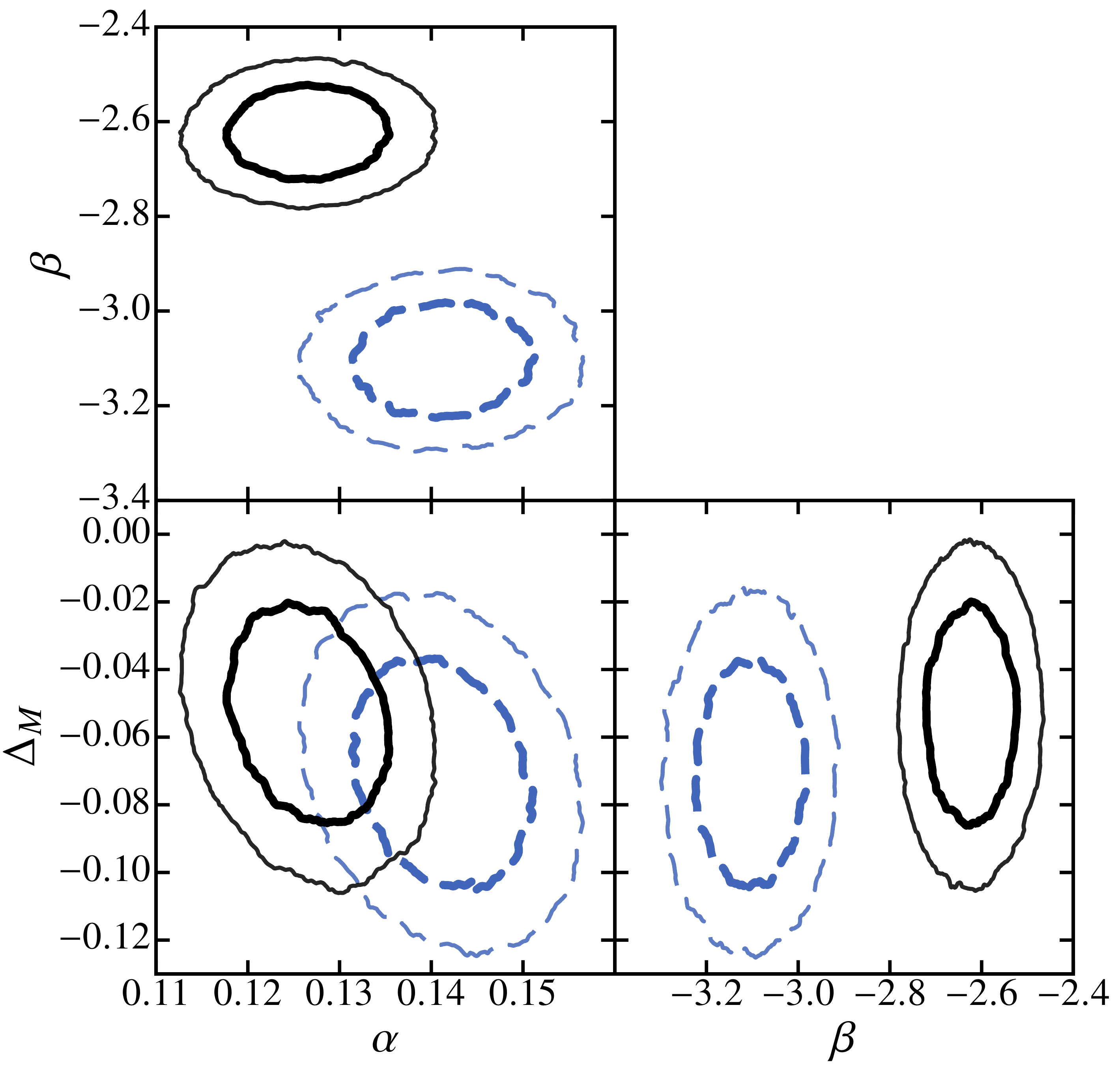}
    \caption{Marginal $0.683$ and $0.95$ credibility levels for pairs of
    standardization parameter derived from the Bayesian (black solid lines)
    and $\chi^2$ (blue dashed lines) analysis.}
    \label{fig:stdparam}
\end{figure}

\begin{table}
\caption{Marginalized mean and standard deviation of SN Ia standardization
    parameters inferred from the full posterior analysis and from the $\chi^2$
    approach for the $w$CDM model, along with the number of $\sigma$ bias.}
\label{tab:wcdmcalib}
\begin{tabular}{cccc}
    \hline
    & This work & $\chi^2$ analysis & Bias amplitude \\
    \hline
    $\alpha$ & $\phantom{-}0.127 \pm 0.006$ & $\phantom{-}0.141 \pm 0.007$ &
    $2\sigma$ \\
    $\beta$ & $-2.62 \pm 0.07$   & $-3.10 \pm 0.08$ & $6\sigma$  \\
    $\Delta_M$ & $-0.053 \pm 0.022$ & $-0.071 \pm 0.023$ & $0.8\sigma$ \\
    \hline
\end{tabular}
\end{table}

\begin{figure}
    \includegraphics[width=.9\columnwidth]{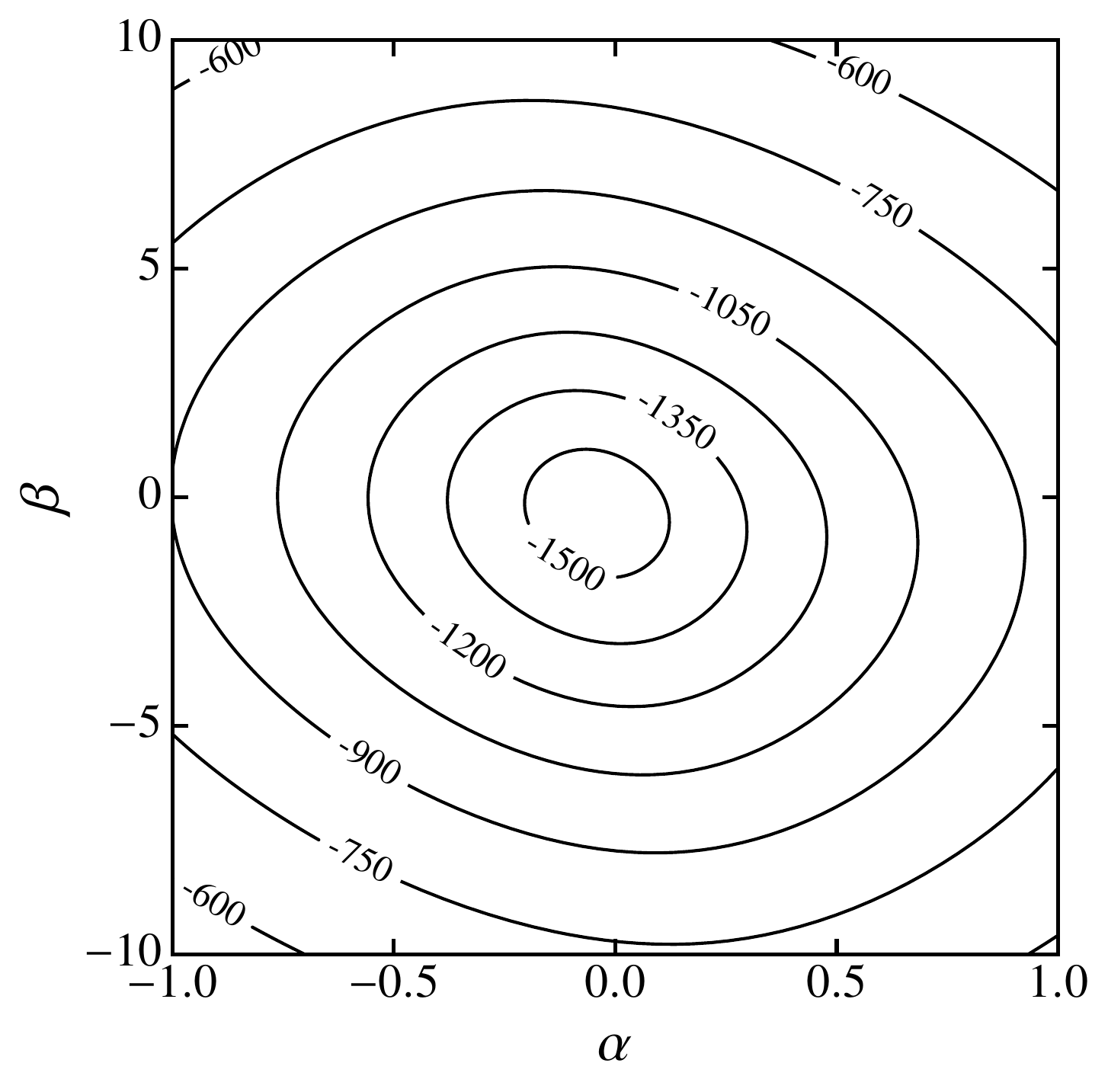}
    \caption{Level contours of $[  \ln \det \mtx{C}_{\text{d}}(\alpha, \beta)
	] / 2$ for the JLA data set. This can be interpreted as the
	log-distortion of the priors on $\alpha$ and $\beta$.}
    \label{fig:distortion}
\end{figure}

In \figref{fig:stdparam} we plot the posterior contours for
different combinations of standardization parameter pairs, while in
\autoref{tab:wcdmcalib} we quote the marginal mean and standard deviation of
$\alpha$, $\beta$ and $\Delta_M$ respectively. We may notice that the values
derived in the $\chi^2$ case are consistent with those quoted in
\citetalias{2014A&A...568A..22B}. We can see that the $\chi^2$ analysis
significantly shifts the standardization parameters away from the ideal
standard candle case (i.e.\ $\alpha = \beta = \Delta_M = 0$) compared to the
Bayesian approach. In particular, we have systematic offsets of $2\sigma$ for
the stretch parameter, $6\sigma$ for the colour correction parameter and about
$1\sigma$ for the host galaxy correction. This indicates that the data
require less adjustment of the light-curve shape, SN colour and host stellar
mass, which is a direct consequence of the fact that neglecting the covariance
term in the $\chi^2$ analysis is equivalent to a distortion of the parameter
priors.  In fact, it amounts to the replacement $\ln P(\vv{\varTheta}) \to \ln
P(\vv{\varTheta}) + \frac{1}{2} \ln \det \mtx{C}_\text{d}(\vv{\varTheta})$ up
to a normalization, thus leading to a level of distortion of the uniform prior
on $\alpha$ and $\beta$ as shown in \figref{fig:distortion}. We can now see
why the $\chi^2$ analysis gives larger values of the standardization
parameter. It effectively uses a prior that artificially underestimates the
region where $(\alpha, \beta)$ is close to zero, while it overestimates the
range where it is large.

In the light of these results, we are tempted to conclude that the Bayesian
approach adds more weight to our belief that SN Ia are standard candles, at
least more than what we are led to believe if we use the customary $\chi^2$
analysis.

To test whether the observed level of bias is cosmological model dependent we
have performed similar analyses for \textLambda{}CDM models with or without
non-zero curvature. The marginal mean and variance of the parameters are quoted
in \autoref{tab:lcdm}. We can see that the bias remains of the same amplitude
for the different cosmological model assumptions.

\begin{table*}
    \caption{Marginal mean and standard deviation of model parameters for
    \textLambda{}CDM models as inferred from the Bayesian method
    of this work and from $\chi^2$ analysis.}
    \label{tab:lcdm}
\begin{tabular}{ccccc}
    \hline
    & \phantom{$-$}\textLambda{}CDM & \phantom{$-$}\textLambda{}CDM ($\chi^2$)
    & \phantom{$-$}Flat \textLambda{}CDM & \phantom{$-$}Flat \textLambda{}CDM
    ($\chi^2$) \\
    \hline
    $\alpha$ & $\phantom{-}0.126 \pm 0.006$ & $\phantom{-}0.141 \pm 0.006$ &
    $\phantom{-}0.126 \pm 0.006$ &  $\phantom{-}0.141 \pm 0.007$ \\
    $\beta$ & $\phantom{0}{-2.62} \pm 0.07\phantom{0}$ & $\phantom{0}{-3.10}
    \pm 0.08\phantom{0}$ & $\phantom{0}{-2.62} \pm 0.07\phantom{0}$ &
    $\phantom{0}{-3.10} \pm 0.08\phantom{0}$ \\
    $\Delta_M$ & $-0.053 \pm 0.022$ & $-0.071 \pm 0.023$ & $-0.053 \pm 0.022$
    & $-0.070 \pm 0.023$ \\
    $\Omega_{\text{M}}$ & $\phantom{-0}0.22 \pm 0.10\phantom{0}$
    & $\phantom{-0}0.20 \pm 0.10\phantom{0}$ & $\phantom{-0}0.33 \pm
    0.03\phantom{0}$ & $\phantom{-0}0.30 \pm 0.03\phantom{0}$ \\
    $\Omega_{\text{DE}}$ & $\phantom{-0}0.50 \pm 0.15\phantom{0}$
    & $\phantom{-0}0.55 \pm 0.15\phantom{0}$ & \phantom{$-0$}N/A\phantom{$0$}
    & \phantom{$-0$}N/A\phantom{$0$} \\
    \hline
\end{tabular}
\end{table*}

\begin{figure}
    \includegraphics[width=.9\columnwidth]{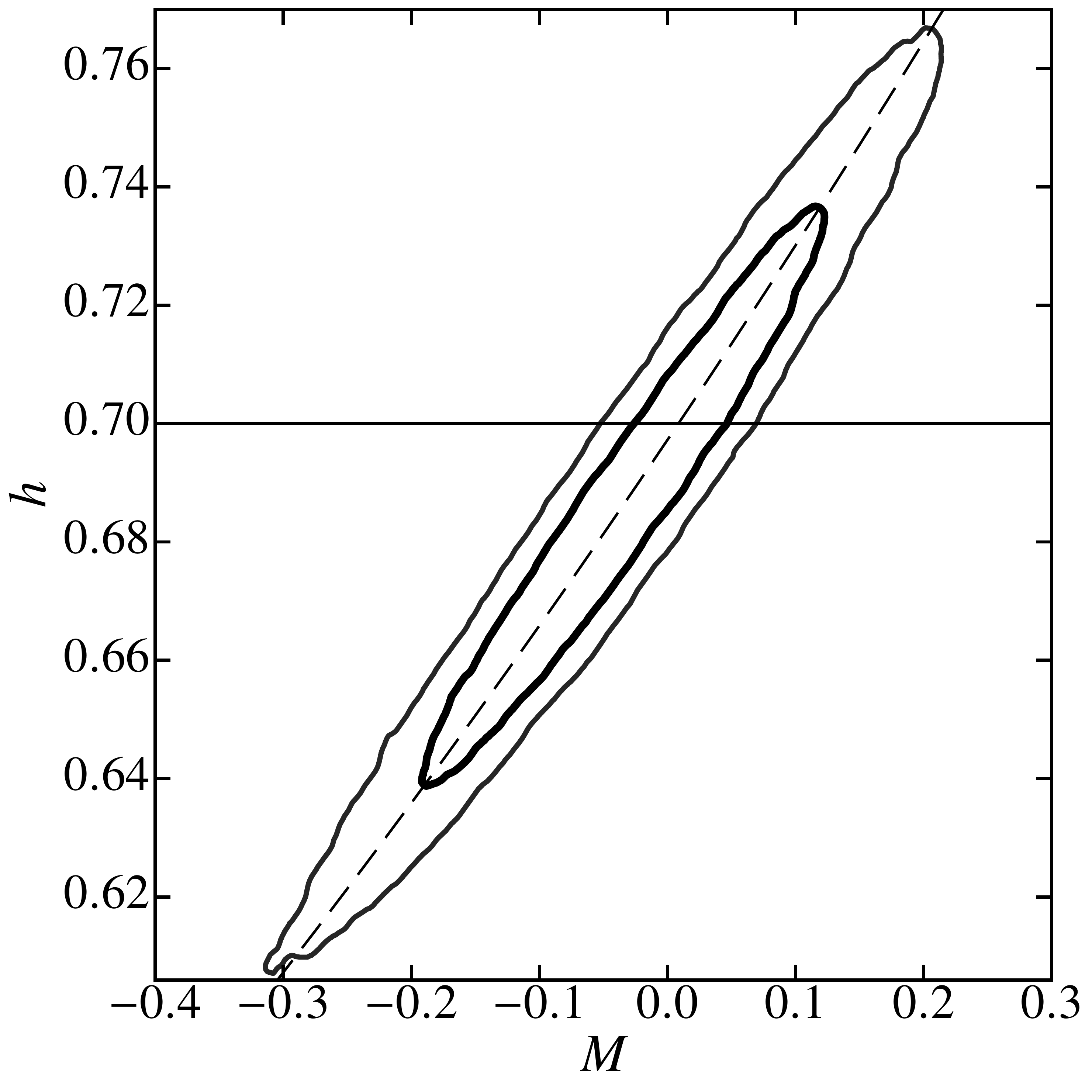}
    \caption{Marginal contours in the $M$--$h$ plane from the full posterior
	analysis for the flat $w$CDM model.  The solid horizontal line
	indicates the value $h = 0.7$ used in
	\citetalias{2014A&A...568A..22B}, while the dashed curve shows the
	degeneracy direction $M - 5 \logg h = \text{const.}$}
     \label{fig:mh}
\end{figure}

Independently of the underlying cosmological model, we find no information gain
on $h$, whose posterior remains indistinguishable from the assumed Gaussian
prior. On the other hand, we find $M = -0.03 \pm 0.11$ for $w$CDM and in the
case of the non-flat \textLambda{}CDM cosmology, while $M = -0.04 \pm 0.11$ for
the flat \textLambda{}CDM case. As expected, the joint posterior in the
$M$--$h$ plane shows a strong degeneracy along the direction $M' = M - 5\logg
h$ as shown in \figref{fig:mh}.  From the marginalized posterior, we find
$\sigma(M) \approx 0.1$.  This reflects the posterior dispersion of $M_B^1$
that should have been there if we had chosen to let it vary freely (see
Sections~\ref{sec:jladesc} and \ref{sec:lumpre}).  If we had neglected $M$
altogether in our model specifications, by degeneracy this could have led to a
spurious constraint on $h$.

\section{Data compression}
\label{sec:datacompress}

\subsection{The issue of scalability}

We now turn to the problem of compressing the JLA data set. The need for
compressed data may respond to specific needs of cosmological analysis. For
instance, tests of the distance-duality relation requires luminosity distance
estimates at redshift locations where angular-diameter distance measurements
are also available. However, the main application of data compression is to
address the problem of scalability. With the increasing size of SN data sets,
the evaluation of expressions such as \eref{eq:fullpost} will be
computationally more challenging. In particular, evaluation of forms such as
$\vv{y} = \mtx{C}_{\text{d}}^{-1} \vv{x}$ by solving the linear system of
equations $\mtx{C}_{\text{d}} \vv{y} = \vv{x}$ will be inefficient when
$\mtx{C}_{\text{d}}$ is parameter dependent.  This is caused by the
computational complexity scaling as $\mathcal{O}(n^3)$, or cubic of the data
size \citep[see e.g.][chapter~4.2.5]{gvl}.  With changing parameter, for
instance during MCMC evaluation, this cubic operation has to be performed
whenever the parameters change value.

The scalability issue of computing a parameter-dependent covariance matrix has
been gaining more attention recently, especially in the context of statistical
data analyses dedicated to measurements of the clustering of matter in the
universe. As an example, \citet{2015JCAP...12..058W} have developed an
interpolation method for efficiently evaluating the likelihood of the two-point
correlation function of the matter density field. A different approach to
handle large data sets with covariances relies instead on approximating the
object of interest with a reduced set of functional bases. This method has seen
widespread use in the context of cosmic microwave background data analysis
\citep*[see][]{1997PhRvD..55.5895T, 1997ApJ...480...22T} and has been applied
in \citetalias{2014A&A...568A..22B} to the JLA data set.

Here, we aim to perform a thorough analysis of the distance modulus data
compression procedure in the context of the Bayesian framework we have
discussed in the previous sections. This will enable us to assess the impact
of model assumptions and more importantly the effects of the
parameter-dependent covariance on the data compression itself.

\subsection{Formalism of linear compression}
\label{sec:formlc}

The goal of the compression is to provide the user with a reduced data set of
distance modulus estimates $\vvdc{\mu}$ together with their covariance matrix
$\mtx{C}_{\mathrm{dc}}$ and the post-compression standardization parameters
$\vvdc{\varphi}$ (correlated with $\vvdc{\mu}$).

In \citetalias{2014A&A...568A..22B}, the linear compression of the JLA data
set is performed by first taking the logarithm of the redshift $z$. This is
because the log-transformation of the redshift makes the
cosmological-dependent part of the signal better linearized (as can be seen in
\figref{fig:visualjla}).  Then, the distance modulus data are fitted against a
parametric model that is represented by `broken line segments' with control
points at fixed log-redshift locations $\{x_1 < x_2 < \cdots < x_m\}$ (in this
section and the next, the symbol $x$ will be used for log-redshift).  The
values of the model parameters at the control points define the fitting
parameters of the compression procedure.  Their (posterior) mean and
covariance give the final compressed data set. 

The parametric fitting model can be cast in the form of a linear combination of
unit sawtooth basis functions $b_i$ defined over an interval $S$ with $m$
control points:
\begin{equation}
    b_i (x) =
    \begin{cases}
	\frac{x - x_{i - 1}}{x_i - x_{i - 1}} & x \in [x_{i - 1}, x_i) \cap S,
	\\
	1 - \frac{x - x_i}{x_{i + 1} - x_i} & x \in [x_i, x_{i + 1}) \cap S,
	\\
	0 & \text{otherwise}.
    \end{cases}
    \label{eq:sawtooth}
\end{equation}
For a data set of size $n$, we can define the $n \times m$ matrix $\mtx{B}$
with elements $B_{ij} = b_j(x_i)$. In the matrix $\mtx{B}$ the $j$th column
gives the image of all data locations under the $j$th basis function, while
the $i$th row contains the mapping of all basis at the same location $x_i$.
If the data locations are sorted, then B is a banded matrix. Using this
definition, the reconstructed data vector can be written as a linear
combination of the basis functions as given by the linear transformation
\begin{equation}
    \vvr{\mu} = \mtx{B} \vv{\xi}
\end{equation}
where the vector $\vv{\xi}$ contains the {\em compression coefficients}. The
goal of the statistical compression analysis is to fit this unknown vector
$\vvr{\mu}$ against the uncompressed data set to determine the coefficients
$\vv{\xi}$.

It is worth noticing that the specific choice of $\mtx{B}$ is more or less
arbitrary.  The form of the basis functions may be dictated by the needs of the
problem at hand. For instance, if the data to be compressed have structures in
the scale space, a set of wavelet bases would be a well-motivated choice
\citep[see e.g.][]{1998ApJ...496....9P}. On the other hand, if the goal is to
extract low-variance, discriminating information from noisy data at the cost of
bias, then the suitable bases may be found through principal component analysis
methods \citep[see][]{2003PhRvL..90c1301H, 2005PhRvD..71b3506H}.

We adopt the sawtooth bases used in \citetalias{2014A&A...568A..22B} which are
especially suitable when the signal to be extracted is expected to be fairly
continuous over the support interval $S$ (as in the case of the distance
modulus). The sawtooth bandwidth is set by the user. A constant value of the
bandwidth corresponds to evenly spaced control points. On the other hand it is
possible to even out statistical noise by adjusting the sawtooth window such as
to cover the same number of data points, a choice that prevents sawtooth
windows to cover insufficient data, which may result in over-fitting.

\subsection{Approximate solution and optimal compression}
\label{sec:compreapprox}

\begin{figure}
    \centering
    \includegraphics[width=.85\columnwidth]{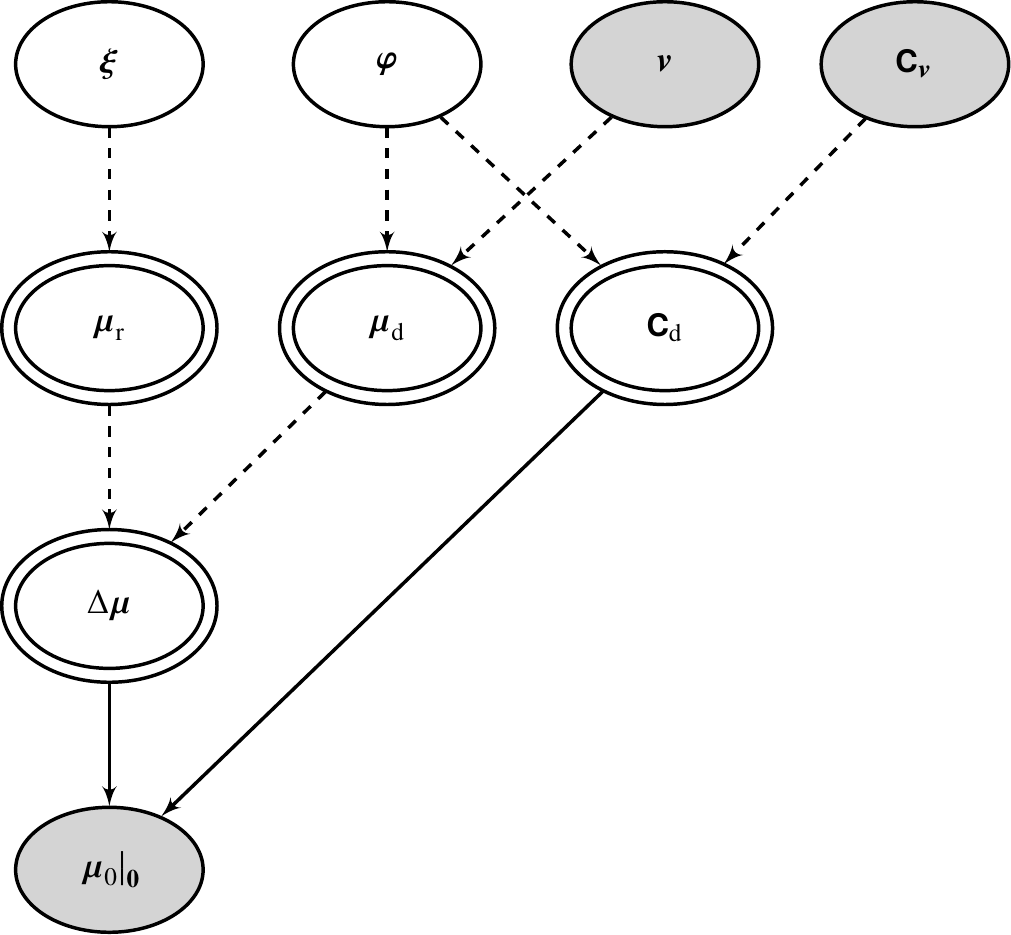}
    \caption{Graphical model for the linear compression of JLA data set.}
    \label{fig:linearc}
\end{figure}

A graphical model for the linear data compression problem of SN data is shown
in \figref{fig:linearc}. The corresponding posterior distribution of the
compression coefficients $\vv{\xi}$ and the standardization parameters
$\vv{\varphi}$ given the uncompressed data set reads as:
\begin{equation}
    \begin{split}
	\ln \pcond{\vv{\varphi}, \vv{\xi}}{\vv{\mu}_0,\ \vv{v},
	\mtx{C}_{\vv{v}}} = &-\ln Z - \frac{n}{2} \ln (2 \upi) - \frac{1}{2}
	\ln \det \mtx{C}_{\text{d}}(\vv{\varphi}) \\
	&- \frac{1}{2} \chi^2(\vv{\varphi}, \vv{\xi}) + \ln P(\vv{\xi}) + \ln
	P(\vv{\varphi}),
    \end{split}
    \label{eq:fpcompres}
\end{equation}
where
\begin{equation}
    \chi^2(\vv{\varphi}, \vv{\xi}) =
    \left[ \vv{\mu}_{\text{d}}(\vv{\varphi}) - \mtx{B} \vv{\xi}
    \right]^{\intercal}
    \mtx{C}_{\text{d}}^{-1}(\vv{\varphi})
    \left[ \vv{\mu}_{\text{d}}(\vv{\varphi}) - \mtx{B} \vv{\xi} \right].
    \label{eq:losschi2}
\end{equation} 
For uniform priors the posterior is globally maximized at an optimal point
$(\vv{\varphi}^{\star}, \vv{\xi}^{\star})$ which maximizes the $\ln P$
function given by \eref{eq:fpcompres}.

The sampling of the posterior can be performed through standard MCMC sampling
as that used in \secref{sec:jlacosmo}. However, the use of a Gaussian
approximation of \eref{eq:fpcompres} can greatly simplify the task. Let
us denote $\vv{\varPhi} = \vv{\varphi} \oplus \vv{\xi}$, then expanding
\eref{eq:fpcompres} about the vector $\vv{\varPhi}^{\star}$ up to second order
about $\vv{\varPhi}^{\star}$, we have
\begin{equation}
    \ln P(\vv{\varPhi}) \approx \ln P(\vv{\varPhi}^{\star}) - \frac{1}{2}
    \vv{D}^{\star} \vv{\cdot} \left(\Delta \vv{\varPhi} \right) -
    \frac{1}{4} \left( \Delta \vv{\varPhi} \right)^{\intercal} \mtx{H}^{\star}
    \left( \Delta \vv{\varPhi} \right),
    \label{eq:taylor}
\end{equation}
where $\Delta \vv{\varPhi} = \vv{\varPhi} - \vv{\varPhi}^{\star}$ and
\begin{equation}
    \vv{D}^{\star} = \left.\frac{\upartial \left( -2 \ln P \right)}{\upartial
    \vv{\varPhi}}\right\vert_{\vv{\varPhi}^{\star}},
    \quad
    \mtx{H}^{\star} = \left.\frac{\upartial^2 \left( -2 \ln P
	    \right)}{\upartial \vv{\varPhi}^2}\right\vert_{\vv{\varPhi}^{\star}}
\end{equation}
are the Jacobian and Hessian of $-2 \ln P$ respectively. Expressions for the
Jacobian and Hessian are straightforward, yet they involve cumbersome algebra
and we do not report them for conciseness. The Jacobians of the terms
proportional to $\vvd{\mu}$ and $\vvr{\mu}$ in \eref{eq:fpcompres} are both
constant, while the derivative of $\ln \det \mtx{C}_{\text{d}}$ is given by
Jacobi's formula
\begin{equation}
    \frac{\upartial \ln \det \mtx{C}}{\upartial \vv{\varPhi}} = \tr \left[
    \mtx{C}^{-1} \frac{\upartial \mtx{C}}{\upartial \vv{\varPhi}} \right].
    \label{eq:jformula}
\end{equation}
To evaluate the Hessian and high-order derivatives, we can iteratively apply
the formula for the derivative of inverse matrix
\begin{equation}
    \frac{\upartial \mtx{C}^{-1}}{\upartial \vv{\varPhi}} = -\mtx{C}^{-1}
    \frac{\upartial \mtx{C}}{\upartial \vv{\varPhi}} \mtx{C}^{-1}
\end{equation}
and \eref{eq:jformula} after evaluating $\upartial \mtx{C}_{\text{d}} /
\upartial \vv{\varPhi}$ by \eref{eq:mudatacov}.  We use these analytical
expressions to numerically determine $\vv{\varPhi}^{\star}$ which maximizes
\eref{eq:taylor}.  This allows us to find the maximum of the approximated
posterior in a stable and efficient manner, while avoiding the pitfalls due to
unstable approximation of the Hessian matrix \citep*[see e.g.][]{DOVI199187}.

To perform the maximization, we use the trust-region Newton-conjugate-gradient
(\textsc{trust-ncg}) algorithm implementation \citep[][chapter~7.1]{nw} from
the \textsc{python} library
\textsc{scipy.optimize}.\footnote{\url{https://scipy.org/}}  Using analytical
expressions for $\vv{D}$ and $\mtx{H}$, it finds the optimal point
$\vv{\varPhi}^{\star}$ in a few seconds on a typical desktop computer and
evaluates the approximated posterior by computing \eref{eq:taylor} at
$\vv{\varPhi}^{\star}$. This is a Gaussian PDF with mean $\vv{\varPhi}^{\star}$
and covariance $\mtx{C}_{\vv{\varPhi}} = 2 \mtx{H}^{-1}(\vv{\varPhi}^{\star})$
from which the marginal distribution for both the compression coefficients
$\vv{\xi}$ and the post-compression standardization parameters $\vvdc{\varphi}$
is obtained. Then, the code uses the optimal compression coefficients and the
covariance to generate series of distance modulus data $\vvdc{\mu}$ at any
given output log-redshift locations $\tilde{x}_i = \logg \tilde{z}_i$
(specified by the user) by computing the Gaussian random vector $\vvdc{\mu} =
\tilde{\mtx{B}} \vv{\xi}$ with mean $\stmean{\vvdc{\mu}} = \tilde{\mtx{B}}
\vv{\xi}^{\star}$ and covariance $\mtx{C}_{\text{dc}} = \tilde{\mtx{B}}
\mtx{C}_{\vv{\xi}} \tilde{\mtx{B}}^{\intercal}$, where the elements of the
`data-generation matrix' $\tilde{\mtx{B}}$ are $\tilde{B}_{ij} =
b_j(\tilde{x}_i)$.\footnote{Information on the post-compression
standardization parameter vector $\vvdc{\varphi}$ can be included in a concise
form by extending the matrix $\tilde{\mtx{B}}$ into a block-diagonal form
$\tilde{\mtx{B}}' = \left( \begin{smallmatrix} \mtx{I} & \mtx{0} \\ \mtx{0} &
\tilde{\mtx{B}} \end{smallmatrix} \right)$ where $\mtx{I}$ is the $3 \times 3$
identity matrix associated with $\vv{\varphi}$.  Thus, after compression the
joint distribution $P(\vvdc{\varphi}, \vvdc{\mu})$ is given by a Gaussian with
mean $\tilde{\mtx{B}}' \vv{\varPhi}^{\star}$ and covariance $\tilde{\mtx{B}}'
\mtx{C}_{\vv{\varPhi}} \tilde{\mtx{B}}'^{\intercal}$.}

There is considerable freedom in the choice of the output redshift locations
$\tilde{z}_i$ or their logarithm $\tilde{x}_i$.  Nevertheless, one should
avoid putting more than two output $\tilde{x}_i$'s between each pair of
adjacent control points that have been specified at the beginning of the
compression procedure (see \secref{sec:formlc}).  In that case, these
compressed output data will not be affine independent, thus providing little
additional information.  Similarly, given a chosen set of $m$ basis functions,
there is no purpose in generating more than $m$ compressed data points,
because the additional points will be inevitably affine dependent on the
others (a consequence of the pigeon-hole principle). A special choice of the
redshift locations is given by setting them to the control points. In such a
case, the data-generation matrix $\tilde{\mtx{B}}$ (or $\tilde{\mtx{B}}'$ for
the inclusion of post-compression standardization parameters) is the identity
matrix $\mtx{I}$, and no actual computation for data generation needs to be
done.

The JLA data compression code we have developed for this analysis is publicly
available.\footnote{\url{https://gitlab.com/congma/libsncompress}} In Appendix
\ref{sec:cdatatab}, we present the result of this compression at the same
redshift locations as those of \citetalias{2014A&A...568A..22B}.

\section{Assessment of compressed data}
\label{sec:jla-comp-anls}

\subsection{Comparison with \citetalias{2014A&A...568A..22B} compression}

Here, we present the results of the JLA data compression analysis. Our goal is
to evaluate the impact of the parameter-dependent covariance on the resulting
data set and compare it with the compressed sample from
\citetalias{2014A&A...568A..22B}.  Following the prescription adopted in
\citetalias{2014A&A...568A..22B}, we set $31$ log-equidistant control points
in the redshift range $z \in [0.01, 1.30]$ for the sawtooth basis defined in
\eref{eq:sawtooth}. Using the compression procedure described in the previous
section, we infer the compression coefficients and the post-compression
standardization parameters using the Gaussian approximation of the posterior
distribution to \eref{eq:fpcompres}. We test the accuracy of this
approximation by MC sampling the full posterior distribution. We will refer to
the former as `Approx.' and the latter as `MC'. To compare to the results of
\citetalias{2014A&A...568A..22B} we perform an analogous estimation in which
we neglect the $\ln \det \mtx{C}_{\text{d}}$ term in \eref{eq:fpcompres},
cases which we will refer to as `Approx.-$\chi^2$' and `MC-$\chi^2$'
respectively.

\begin{figure}
    \includegraphics[width=.9\columnwidth]{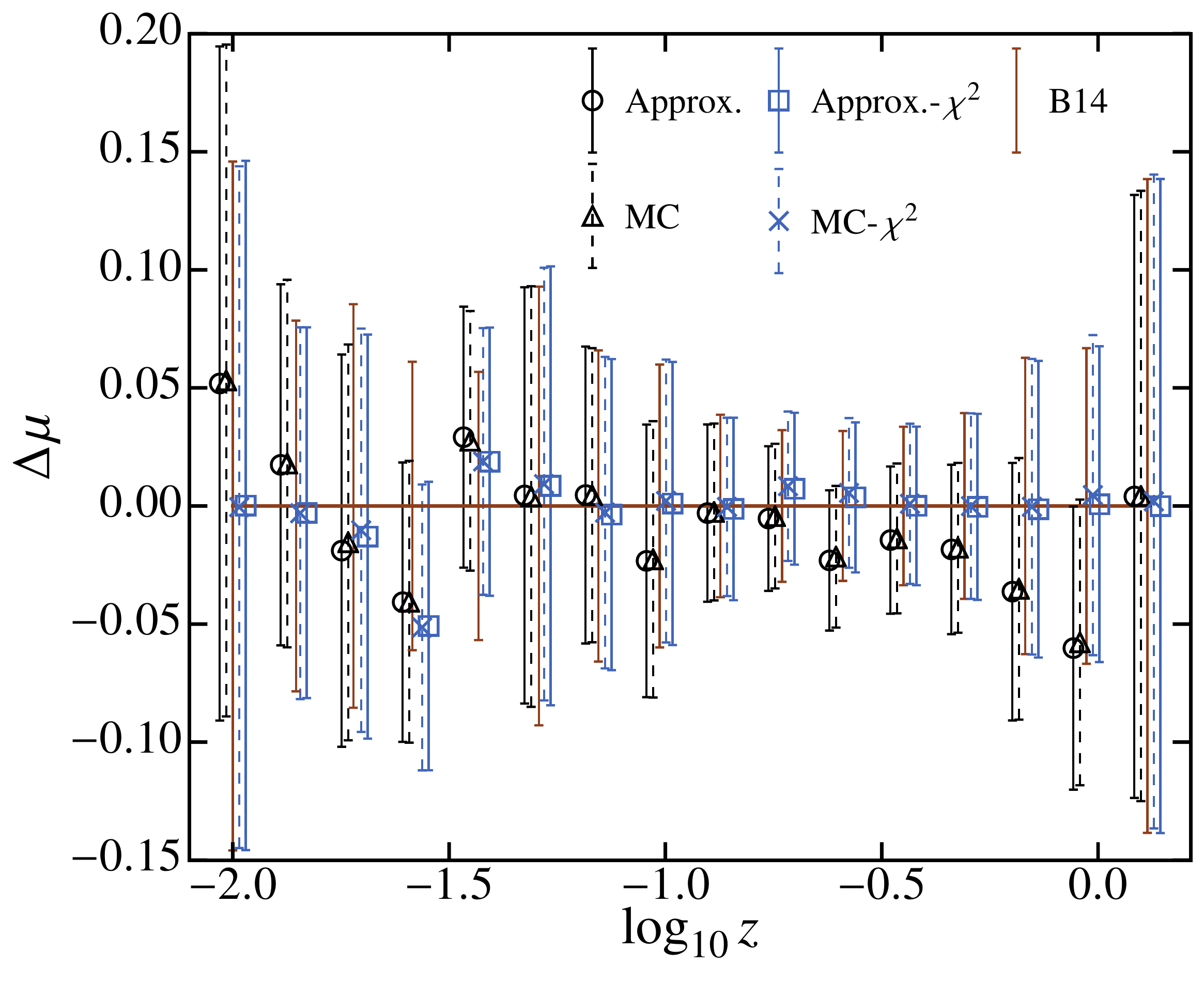}
    \caption{Difference of mean compressed distance moduli with respect to the
	compressed data set of \citetalias{2014A&A...568A..22B}, obtained from
	the Gaussian approximation of the posterior (black circles), MC
	computation of the posterior (black triangles), quadratic
	approximation of the $\chi^2$ (blue squares) and MC sampling of the
	$\chi^2$ (blue crosses).  For visual purposes, we only display one
	every two data points staggered around the redshift locations by
	0.015~dex to reduce crowding in the figure.  The error bars show the
	size of the marginalized deviations corresponding to each control
    point from the respective data set.}
    \label{fig:compdata}
\end{figure}

In \figref{fig:compdata} we plot the deviations of the mean of the generated
distance moduli with respect to the compressed data from table F.1 of
\citetalias{2014A&A...568A..22B} obtained from the Gaussian approximation of
the posterior, the MC computation of the posterior, the quadratic
approximation of the $\chi^2$, and the MC sampling of the $\chi^2$.  The error
bars are the marginalized standard deviations at each control point.  They are
displayed only for a qualitative visual comparison, since the figure
does not reflect the full covariance of the compressed data which we will
discuss later.

We can see that the results from the use of the Gaussian approximation are
indistinguishable from those obtained using the MC sampling even in the case
where the $\ln \det \mtx{C}_{\text{d}}$ term is neglected. This also
guarantees that our optimization algorithm for the determination of the
parameter vector $\vv{\varPhi}^{\star}$ has converged to a global maximum
(minimum for the $\chi^2$ analysis) instead of a local one.

Let us now compare the differences of the generated distance moduli to those
from \citetalias{2014A&A...568A..22B}. The latter are consistent with the
compression obtained from the $\chi^2$ analysis for which differences are
below $0.1$~mag and well within $1\sigma$ errors especially at $z \ge 0.1$
where differences vanish. However, we can also notice that the
\citetalias{2014A&A...568A..22B} compressed data set shows deviations as large
as $1\sigma$ with respect to the result of the Bayesian analysis.

As for the standardization parameters, in \autoref{tab:compcalib} we quote
their marginal mean and standard deviation, post-compression, obtained using
different methods. Again, we can see that the estimates from the Gaussian
approximation are consistent with the MC results and in agreement with the
values inferred from the full data set shown in Tables~\ref{tab:wcdmcalib} and
\ref{tab:lcdm}.

\begin{table*}
  \caption{Marginal mean and standard deviation of standardization parameters
  after compression for the various cases.}
\begin{tabular}{ccccc}
    \hline
    & \phantom{$-$}Approx. & \phantom{$-$}Approx.-$\chi^2$ & \phantom{$-$}MC &
    \phantom{$-$}MC-$\chi^2$ \\
    \hline
    $\alpha$ & $\phantom{-}0.125 \pm 0.006$ & $\phantom{-}0.140 \pm 0.007$ &
    $\phantom{-}0.126 \pm 0.007$ & $\phantom{-}0.141 \pm 0.006$ \\
    $\beta$ & $\phantom{0}{-2.58} \pm 0.07\phantom{0}$ & $\phantom{0}{-3.08}
    \pm 0.08\phantom{0}$ & $\phantom{0}{-2.60} \pm 0.08\phantom{0}$ &
    $\phantom{0}{-3.11} \pm 0.07\phantom{0}$ \\
    $\Delta_M$ & $-0.052 \pm 0.022$ & $-0.070 \pm 0.023$ & $-0.053 \pm 0.023$
    & $-0.071 \pm 0.022$ \\
    \hline
\end{tabular}
\label{tab:compcalib}
\end{table*}

In order to quantify differences between the estimated covariance matrices we
consider two diagnostics.  The first is the ratio of matrix determinant scaled
by the number of parameters $m$,
\begin{equation}
    \label{eq:uncperd}
    r = \left( \frac{\det \mtx{C}_2}{\det \mtx{C}_1} \right)^{\frac{1}{2 m}},
\end{equation}
which quantifies by which factor the Gaussian uncertainties scale up from
$\mathcal{N}_1$ to $\mathcal{N}_2$, per dimension.  The second is the
Kullback--Leibler (KL) divergence \citep{kl} from random variable $P$ to $Q$,
defined as
\begin{equation}
    \label{eq:defkl}
    D_{\text{KL}}(P \, \Vert \, Q) = \int \ln \left( \frac{\ud P}{\ud Q}
    \right) \ud P.
\end{equation}
As we are interested in differences between covariances, we compute the KL
divergence by shifting the mean of one of the distributions to coincide with
the other.  In this case, for our compressed SN Ia data with Gaussian
approximation, it is a function of the covariance matrices:
\begin{equation}
    \label{eq:klgauss}
    D_{\text{KL}}(\mathcal{N}_1 \, \Vert \, \mathcal{N}_2) = \frac{1}{2}
    \left[ \ln \left( \frac{\det \mtx{C}_2}{\det \mtx{C}_1} \right) + \tr
    \left( \mtx{C}_2^{-1} \mtx{C}_1 \right) - m \right].
\end{equation}
The $r$ diagnostic in \eref{eq:uncperd} is only a measure of the total
`size' of the uncertainty, while the (centred) KL divergence is a much more
sensitive diagnostic, because \eref{eq:klgauss} is zero if and only if the two
distributions are identical (up to a translation).  It is also sensitive to
the difference in the `shape' or pattern of correlation.

\begin{figure*}
    \includegraphics[width=.9\textwidth]{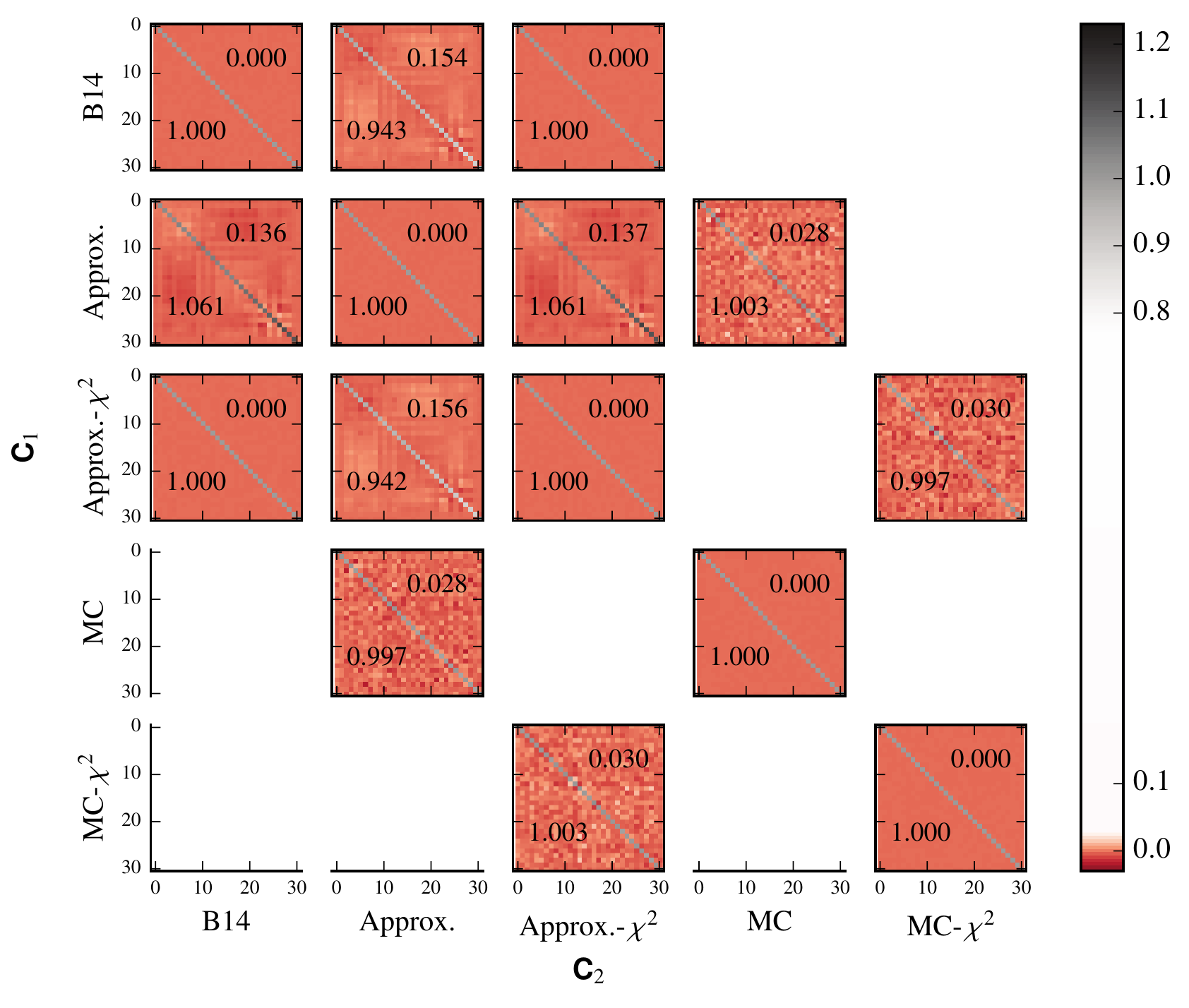}
    \caption{Pairwise comparisons of covariance matrices for compressed
	distance moduli obtained from various computations.  In each panel,
	the comparison matrix $\mtx{W}_{12}$ is displayed as a bitmap image.
	The scaled ratio of the determinants and the centred KL divergence
	values are quoted in the lower-left and upper-right side of each panel
	respectively.}
    \label{fig:diffmatrices}
\end{figure*}

To visualize the differences between pairs of covariances, we introduce an
algebraic method described in Appendix~\ref{apd:covmethod}. This is based
on the idea that two covariance matrices $\mtx{C}_1$ and $\mtx{C}_2$ can be
linked by the matrix $\mtx{W}_{12}$, displayed as a bitmap image. If
$\mtx{C}_1 \approx \mtx{C}_2$, then the matrix $\mtx{W}_{12}$ is close to the
identity.  If they differ by a simple scaling, then $\mtx{W}_{12}$ is diagonal
with diagonal elements differing from unity.  On the other hand, if
differences occur on off-diagonal elements these will stand out as
off-diagonal features on the image of $\mtx{W}_{12}$. In
\figref{fig:diffmatrices}, we display $\mtx{W}_{12}$ between pairs of matrices
for the different cases. In each panel, we also quote the ratio $r$ and the
centred KL divergence value $D_{\text{KL}}$. Comparison between some of the
pairs is not shown since it would only provide redundant information.  We use
a colour palette suitable for the bimodal distribution of all pixel values.

First, comparing $\mtx{C}_1$ obtained from the Gaussian approximation
of the posterior (`Approx.') to $\mtx{C}_2$ from MC sampling of the full
posterior distribution (`MC'), we can see they are nearly identical with
differences in the off-diagonal elements simply due to MC noise, an artefact
of numerical computation.  This is also confirmed quantitatively by the
vanishing $D_{\text{KL}} \approx 0.03$ and a negligible difference of $r$ from
unity by 0.3 per cent.

Similarly, we find the covariance $\mtx{C}_1$ of the compressed data from
\citetalias{2014A&A...568A..22B} to be identical to $\mtx{C}_2$ from the
Approx.-$\chi^2$ computation. This is not surprising, since the data
compression performed in \citetalias{2014A&A...568A..22B} neglects the $\ln
\det \mtx{C}_{\text{d}}$ term.  Indeed, neglecting this term leads to
significant, systematic differences between the covariance inferred from the
$\chi^2$ analysis and that obtained from the posterior computation.  The $r$
value indicates that the $\chi^2$ method overstates the overall uncertainty by
6 per cent. The non-vanishing value of $D_{\text{KL}} \approx 0.15$, five
times the noise-induced value, cannot be dismissed as a small random error.
This is visually corroborated by the presence of block structures in the
$\mtx{W}_{12}$ comparison matrix, a feature distinguished from mere numerical
artefacts.

\subsection{Standard-candle properties and cosmological constraints}
\label{sec:stddiscrep}

Here, we use the set of compressed data to perform a consistency analysis of
the SN Ia light-curve parameters across different redshift intervals. Recently,
\citet{2016MNRAS.460.2586L} have performed an analysis of the redshift
evolution of the standardization parameters by dividing the JLA data set in
redshift subsamples and found that the higher redshift data favour a lower
value of colour correction parameter $\beta$ than the subsample at lower
redshift.  We show how the use of compressed data generated from $\chi^2$
analysis may lead to unexpected results when performing such tests.

We divide the JLA data set into two overlapping redshift regions: $S_1$
containing $166$ data points in the redshift range $0.01 \le z < 0.114$ and
$S_2$ containing $599$ data points in the range $0.082 \le z < 1.3$. $S_1$ is
dominated by low-$z$ sources from various observational programmes, while $S_2$
is dominated by SDSS and SNLS sources. The overlapping region covers the
redshift range $0.082 \le z < 0.114$ and contains $25$ points. We apply the
compression to both subsamples with control points at the same locations as
described in the previous sections which is consistent with the data binning of
\citetalias{2014A&A...568A..22B}.  In the overlapping region we find the
distance moduli to be consistent within $1\sigma$. This is an important
consistency check that validates our compression procedure.

\begin{figure}
    \includegraphics[width=\columnwidth]{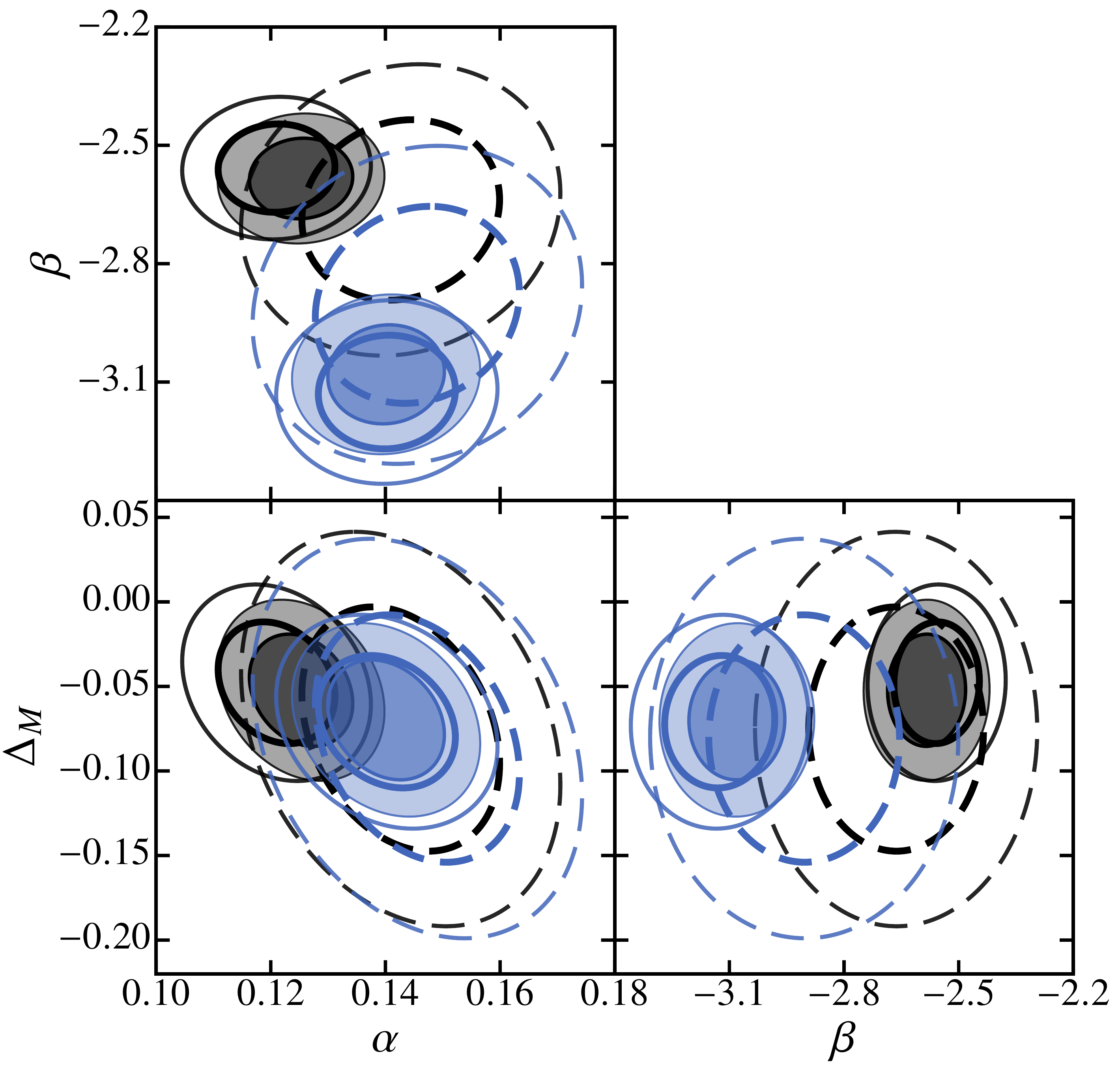}
    \caption{Marginal $0.683$ and $0.95$ contours of the post-compression
	standardization parameters inferred from the Gaussian approximation of
	the posterior (black) and the $\chi^2$ analysis (blue), in $S_1$
	(dashed lines), $S_2$ (solid lines) and the full data set (filled
	contours), respectively.}
    \label{fig:ccombine}
\end{figure}

In \figref{fig:ccombine} we plot the contours of the marginalized
Gaussian-approximate PDF for the post-compression standardization parameters
inferred from the Bayesian analysis with Gaussian approximation of the
posterior and the $\chi^2$ approach, in $S_1$, $S_2$, and for the full data
set respectively. We may notice that the constraints obtained using the full
compressed data set are dominated by data in the region $S_2$. This is not
surprising since this redshift interval has the greatest number of data
points.  The ellipses from $S_1$ and $S_2$ intervals lie within $1\sigma$. In
contrast, we can see that the results inferred from the $\chi^2$ computation
favour values of the parameter $\beta$ that are systematically larger (in
absolute value) than those inferred from the posterior analysis.  Again, such
a systematic bias is the result of neglecting the $\ln \det
\mtx{C}_{\text{d}}$ term.

\begin{figure}
    \includegraphics[width=.9\columnwidth]{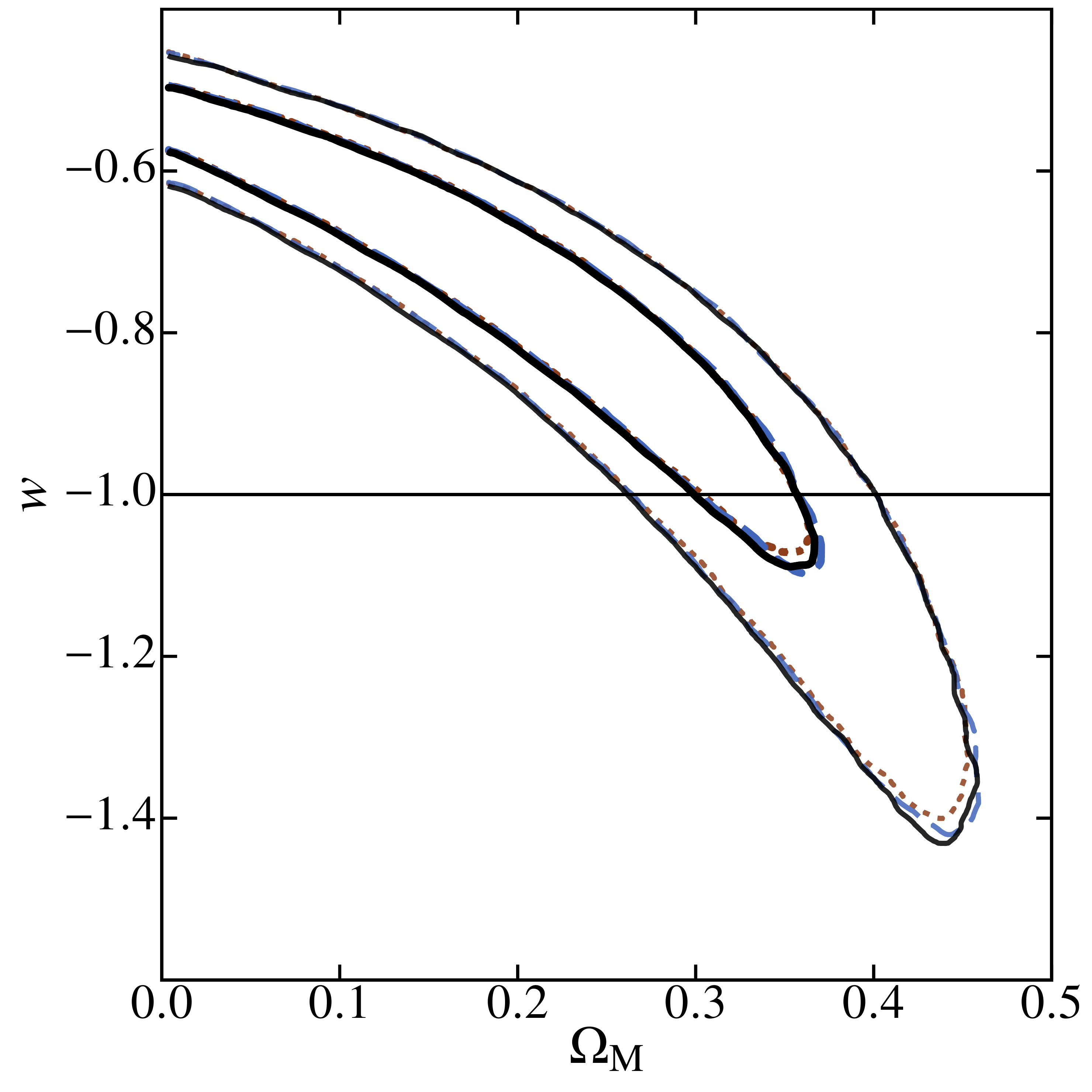}
    \caption{Marginal contours in the $\Omega_{\mathrm{M}}$--$w$ plane from
	the analysis of the full uncompressed JLA data set (black solid lines),
	the compressed set obtained from the Gaussian approximation of the
	posterior (blue dashed lines) and with standardization parameters
	pre-marginalized after the compression procedure (brown dotted lines).}
    \label{fig:wcdmcomp}
\end{figure}

As final test of the data compression analysis, we perform a cosmological
parameter inference using the compressed JLA data in the case of the $w$CDM
model discussed in \secref{sec:jlacosmo}. In \figref{fig:wcdmcomp}, we plot
the contours in the $\Omega_{\text{M}}$--$w$ plane obtained using the
uncompressed full JLA data set, the compressed data from the Gaussian
approximation of the posterior (including the post-compression
standardization parameters) and the compressed data with standardization
parameters pre-marginalized before entering the cosmological fitting.  The
displayed contours are nearly indistinguishable from one another.  Similarly,
we find identical marginal mean and standard deviation of the model
parameters: $w = -0.82 \pm 0.22$ and $\Omega_{\text{M}} = 0.22 \pm 0.11$.
These results are in excellent agreement with those discussed in
\secref{sec:jlacosmo}.  Furthermore, for $(w, \Omega_{\text{M}})$, we estimate
the KL divergence of their two-dimensional distributions, from the one
obtained using compressed data, to the other obtained with the full JLA data,
using the $k$-nearest neighbour estimator of \citet{klknn}.  The resulting
$D_{\text{KL}} = 0.004$ indicates that the cosmological information is
preserved by the data compression model described in
\secref{sec:datacompress}.  To put this minuscule value into context, the
systematic shift from the $\chi^2$ result to the Bayesian posterior to that we
see in \secref{sec:jlacosmo} and \figref{fig:wcdm} corresponds to
$D_{\text{KL}} = 0.51$.

We make publicly available example programs that implement the graphical
models and the MCMC analyses of the full JLA data set and the compressed one
at \url{https://gitlab.com/congma/sn-bayesian-model-example/}.

\section{Conclusion}
\label{sec:conc}

We have performed a detailed Bayesian statistical analysis of the JLA data set
using Bayesian graphical models to derive the full posterior distribution of
fitting model parameters. We have done so with the specific intent of
evaluating the impact of correctly propagating SN standard-candle parameter
errors through the data covariance matrix in contrast to the $\chi^2$
analysis.

Comparing results from the full posterior distribution with those inferred
from the $\chi^2$ approach we find a statistically significant shift of the SN
standard-candle corrections towards lower (absolute value).  This is because
the $\chi^2$ fit does not fully propagate the parameter dependence of the
covariance which contribute with a $\ln \det \mtx{C}_{\mathrm{d}}$ term in the
parameter posterior. We have shown that neglecting this term is equivalent to
assuming non-uniform priors on the parameter $\alpha$ and $\beta$ which
parametrize the effect of the SN light-curve stretch and colour in the
standard-candle relation. Due to this improper prior, the $\chi^2$ analysis
gives more statistical weight to the region of the parameter space away from
$\alpha = \beta = 0$. In particular, we find a $2\sigma$ shift in the
best-fitting value of $\alpha$, a $6\sigma$ change in the best-fitting value
of $\beta$ and lower host galaxy correction $\Delta_M$ of roughly $1\sigma$.
\citet{2010MNRAS.406..782S} found a non-vanishing $\Delta_M$ at $3.7\sigma$.
\citetalias{2014A&A...568A..22B} measured a non-zero value at $5\sigma$
(excluding the systematics of the host mass correction itself) or $3\sigma$
(including all systematics), while our estimate is at $2.4\sigma$.  Recently,
\citet*{2016MNRAS.457.3470C} also reported a $2.5\sigma$ difference based on
the same host mass classification in the SDSS-II SN Ia.  We find the amplitude
of the systematic offset between the full Bayesian analysis and the $\chi^2$
results to be independent of the underlying cosmological model assumption.

The impact of the $\chi^2$ analysis bias is less significant on the
cosmological parameter inference. To this purpose we have derived marginal
bounds on the parameters of a flat $w$CDM. The constraints on
$(\Omega_{\mathrm{M}}, w)$ from the two statistical approaches differ to within
$\sim 1\sigma$. However, the effect can be more significant if the bounds are
combined with other constraints that break the cosmological degeneracies of
the distance modulus. 

This statistical bias problem also affects the generation of
compressed distance modulus data. We have used the linear compression model
presented in \citetalias{2014A&A...568A..22B} and determined the compression
coefficients performing a full posterior analysis of the compression
parameters and post-compression standardization parameters as opposite to the
$\chi^2$ approach. Indeed, the comparison between the compressed data sets
obtained using the full posterior analysis and the $\chi^2$ approach shows
differences of the marginal mean value of the post-standardization parameters,
the mean of the compressed distance moduli, and their covariance.

In related works dedicated to SN Ia cosmology with Bayesian methods
\citep{2011MNRAS.418.2308M, 2014MNRAS.437.3298M, 2015ApJ...813..137R,
2016ApJ...827....1S}, cosmological models are analysed globally with the SN Ia
observables.  Although we acknowledge that these analyses are better equipped
with representing the full dependence relations of all the random variables
involved, we also note the considerable cost and complexity of such analyses.
In this work, we instead take the already reduced SALT2 filter data output of
JLA as statistical evidence (and we expect that future data may be utilized in
a similar manner).  This allows us to present a simple, modular approach of
using the reduced data for a wide family of cosmological models.  The
simplicity is further improved by the data compression procedure.  This step
is present in \citetalias{2014A&A...568A..22B} but lacking formal details.  In
this work, we formalize the compression as a discrete linear model and subject
it to the same Bayesian analysis showing that inconsistency exists in the
\citetalias{2014A&A...568A..22B} compression results.

Our main contribution to the Bayesian data compression problem of the JLA
data set is the development of an efficient method that uses a Gaussian
approximation of the posterior which we have checked against MC sampling. We
have implemented this method as publicly available code that allows the user
to fast generate compressed distance modulus data set (including their
correlated standardization parameters) at given input redshift locations.

The cosmological parameter inference from the compressed data set gives
results that are nearly identical to those obtained using the entire
uncompressed JLA data set, and this shows that cosmological information is
left unaltered by the statistical compression method.  However, we acknowledge
that further investigation should be needed for understanding the extent of
its limitations, and for its greater optimization and generalization towards
future data sets.

Overall, the analysis presented here stresses the necessity of using
self-consistent Bayesian statistical approaches to perform unbiased model
parameter inference of SN Ia to generated unbiased sets of compressed data.

\section*{Acknowledgements}

The research leading to these results has received funding from the European
Research Council under the European Union's Seventh Framework Programme
(FP/2007--2013)/ERC Grant Agreement no.~279954.  CM acknowledges the support
from the joint research training programme of the Chinese Academy of Sciences
and the CNRS.  Data visualizations are prepared with the
\textsc{\mbox{matplotlib}}\footnote{\url{http://matplotlib.org/}} library
\citep{Hunter07}, and DAG diagrams are drawn using the
\textsc{dot2tex}\footnote{\url{https://dot2tex.readthedocs.org/}} tool created
by K.~M.~Fauske and contributors.  The authors are grateful for the critical,
anonymous reviews that improved this paper.

\bibliographystyle{mnras}
\bibliography{ms}

\begin{thebibliography}{}
\makeatletter
\relax
\def\mn@urlcharsother{\let\do\@makeother \do\$\do\&\do\#\do\^\do\_\do\%\do\~}
\def\mn@doi{\begingroup\mn@urlcharsother \@ifnextchar [ {\mn@doi@}
  {\mn@doi@[]}}
\def\mn@doi@[#1]#2{\def\@tempa{#1}\ifx\@tempa\@empty \href
  {http://dx.doi.org/#2} {doi:#2}\else \href {http://dx.doi.org/#2} {#1}\fi
  \endgroup}
\def\mn@eprint#1#2{\mn@eprint@#1:#2::\@nil}
\def\mn@eprint@arXiv#1{\href {http://arxiv.org/abs/#1} {{\tt arXiv:#1}}}
\def\mn@eprint@dblp#1{\href {http://dblp.uni-trier.de/rec/bibtex/#1.xml}
  {dblp:#1}}
\def\mn@eprint@#1:#2:#3:#4\@nil{\def\@tempa {#1}\def\@tempb {#2}\def\@tempc
  {#3}\ifx \@tempc \@empty \let \@tempc \@tempb \let \@tempb \@tempa \fi \ifx
  \@tempb \@empty \def\@tempb {arXiv}\fi \@ifundefined
  {mn@eprint@\@tempb}{\@tempb:\@tempc}{\expandafter \expandafter \csname
  mn@eprint@\@tempb\endcsname \expandafter{\@tempc}}}

\bibitem[\protect\citeauthoryear{{Astier} et~al.,}{{Astier}
  et~al.}{2006}]{2006A&A...447...31A}
{Astier} P.,  et~al., 2006, \mn@doi [\aap] {10.1051/0004-6361:20054185}, \href
  {http://adsabs.harvard.edu/abs/2006A%26A...447...31A} {447, 31}

\bibitem[\protect\citeauthoryear{{Bassett}, {Corasaniti}  \& {Kunz}}{{Bassett}
  et~al.}{2004}]{2004ApJ...617L...1B}
{Bassett} B.~A.,  {Corasaniti} P.-S.,   {Kunz} M.,  2004, \mn@doi [\apj]
  {10.1086/427023}, \href {http://adsabs.harvard.edu/abs/2004ApJ...617L...1B}
  {617, L1}

\bibitem[\protect\citeauthoryear{{Betoule} et~al.,}{{Betoule}
  et~al.}{2014}]{2014A&A...568A..22B}
{Betoule} M.,  et~al., 2014, \mn@doi [\aap] {10.1051/0004-6361/201423413},
  \href {http://adsabs.harvard.edu/abs/2014A%26A...568A..22B} {568, A22}

\bibitem[\protect\citeauthoryear{{Brooks} \& {Gelman}}{{Brooks} \&
  {Gelman}}{1998}]{brooksgelman}
{Brooks} S.~P.,  {Gelman} A.,  1998, \mn@doi [J.\ Comput.\ Graph.\ Stat.]
  {10.1080/10618600.1998.10474787}, 7, 434

\bibitem[\protect\citeauthoryear{{Campbell} et~al.,}{{Campbell}
  et~al.}{2013}]{2013ApJ...763...88C}
{Campbell} H.,  et~al., 2013, \mn@doi [\apj] {10.1088/0004-637X/763/2/88},
  \href {http://adsabs.harvard.edu/abs/2013ApJ...763...88C} {763, 88}

\bibitem[\protect\citeauthoryear{{Campbell}, {Fraser}  \& {Gilmore}}{{Campbell}
  et~al.}{2016}]{2016MNRAS.457.3470C}
{Campbell} H.,  {Fraser} M.,   {Gilmore} G.,  2016, \mn@doi [\mnras]
  {10.1093/mnras/stw115}, \href
  {http://adsabs.harvard.edu/abs/2016MNRAS.457.3470C} {457, 3470}

\bibitem[\protect\citeauthoryear{{Conley} et~al.,}{{Conley}
  et~al.}{2011}]{2011ApJS..192....1C}
{Conley} A.,  et~al., 2011, \mn@doi [\apjs] {10.1088/0067-0049/192/1/1}, \href
  {http://adsabs.harvard.edu/abs/2011ApJS..192....1C} {192, 1}

\bibitem[\protect\citeauthoryear{{Contreras} et~al.,}{{Contreras}
  et~al.}{2010}]{2010AJ....139..519C}
{Contreras} C.,  et~al., 2010, \mn@doi [\aj] {10.1088/0004-6256/139/2/519},
  \href {http://adsabs.harvard.edu/abs/2010AJ....139..519C} {139, 519}

\bibitem[\protect\citeauthoryear{{D'Agostini}}{{D'Agostini}}{2005}]{2005physics..11182D}
{D'Agostini} G.,  2005, preprint, \href
  {http://adsabs.harvard.edu/abs/2005physics..11182D} {} (\mn@eprint {arXiv}
  {physics/0511182})

\bibitem[\protect\citeauthoryear{Dov{\` i}, {Paladino}  \& {Reverberi}}{Dov{\`
  i} et~al.}{1991}]{DOVI199187}
Dov{\` i} V.~G.,  {Paladino} O.,   {Reverberi} A.~P.,  1991, \mn@doi [Appl.\
  Math.\ Lett.] {10.1016/0893-9659(91)90129-J}, 4, 87

\bibitem[\protect\citeauthoryear{{Efstathiou}}{{Efstathiou}}{2014}]{2014MNRAS.440.1138E}
{Efstathiou} G.,  2014, \mn@doi [\mnras] {10.1093/mnras/stu278}, \href
  {http://adsabs.harvard.edu/abs/2014MNRAS.440.1138E} {440, 1138}

\bibitem[\protect\citeauthoryear{{Frieman} et~al.,}{{Frieman}
  et~al.}{2008}]{2008AJ....135..338F}
{Frieman} J.~A.,  et~al., 2008, \mn@doi [\aj] {10.1088/0004-6256/135/1/338},
  \href {http://adsabs.harvard.edu/abs/2008AJ....135..338F} {135, 338}

\bibitem[\protect\citeauthoryear{{Gallagher}, {Garnavich}, {Berlind},
  {Challis}, {Jha}  \& {Kirshner}}{{Gallagher}
  et~al.}{2005}]{2005ApJ...634..210G}
{Gallagher} J.~S.,  {Garnavich} P.~M.,  {Berlind} P.,  {Challis} P.,  {Jha} S.,
    {Kirshner} R.~P.,  2005, \mn@doi [\apj] {10.1086/491664}, \href
  {http://adsabs.harvard.edu/abs/2005ApJ...634..210G} {634, 210}

\bibitem[\protect\citeauthoryear{{Gelman} \& {Rubin}}{{Gelman} \&
  {Rubin}}{1992}]{gelmanrubin}
{Gelman} A.,  {Rubin} D.~B.,  1992, \mn@doi [Stat.\ Sci.]
  {10.1214/ss/1177011136}, 7, 457

\bibitem[\protect\citeauthoryear{{Golub} \& van Loan}{{Golub} \& van
  Loan}{2013}]{gvl}
{Golub} G.~H.,  van Loan C.~F.,  2013, Matrix Computations, 4th edn.
Johns Hopkins Univ.\ Press, Baltimore

\bibitem[\protect\citeauthoryear{{Guy} et~al.,}{{Guy}
  et~al.}{2007}]{2007A&A...466...11G}
{Guy} J.,  et~al., 2007, \mn@doi [\aap] {10.1051/0004-6361:20066930}, \href
  {http://adsabs.harvard.edu/abs/2007A%26A...466...11G} {466, 11}

\bibitem[\protect\citeauthoryear{{Hamilton} \& {Tegmark}}{{Hamilton} \&
  {Tegmark}}{2000}]{2000MNRAS.312..285H}
{Hamilton} A.~J.~S.,  {Tegmark} M.,  2000, \mn@doi [\mnras]
  {10.1046/j.1365-8711.2000.03074.x}, \href
  {http://adsabs.harvard.edu/abs/2000MNRAS.312..285H} {312, 285}

\bibitem[\protect\citeauthoryear{{Hamuy}, {Phillips}, {Suntzeff}, {Schommer},
  {Maza}  \& {Aviles}}{{Hamuy} et~al.}{1996}]{1996AJ....112.2391H}
{Hamuy} M.,  {Phillips} M.~M.,  {Suntzeff} N.~B.,  {Schommer} R.~A.,  {Maza}
  J.,   {Aviles} R.,  1996, \mn@doi [\aj] {10.1086/118190}, \href
  {http://adsabs.harvard.edu/abs/1996AJ....112.2391H} {112, 2391}

\bibitem[\protect\citeauthoryear{{Hicken} et~al.,}{{Hicken}
  et~al.}{2009}]{2009ApJ...700..331H}
{Hicken} M.,  et~al., 2009, \mn@doi [\apj] {10.1088/0004-637X/700/1/331}, \href
  {http://adsabs.harvard.edu/abs/2009ApJ...700..331H} {700, 331}

\bibitem[\protect\citeauthoryear{{Hogg}}{{Hogg}}{1999}]{1999astro.ph..5116H}
{Hogg} D.~W.,  1999, preprint, \href
  {http://adsabs.harvard.edu/abs/1999astro.ph..5116H} {} (\mn@eprint {}
  {astro-ph/9905116})

\bibitem[\protect\citeauthoryear{{Humphreys}, {Reid}, {Moran}, {Greenhill}  \&
  {Argon}}{{Humphreys} et~al.}{2013}]{2013ApJ...775...13H}
{Humphreys} E.~M.~L.,  {Reid} M.~J.,  {Moran} J.~M.,  {Greenhill} L.~J.,
  {Argon} A.~L.,  2013, \mn@doi [\apj] {10.1088/0004-637X/775/1/13}, \href
  {http://adsabs.harvard.edu/abs/2013ApJ...775...13H} {775, 13}

\bibitem[\protect\citeauthoryear{{Hunter}}{{Hunter}}{2007}]{Hunter07}
{Hunter} J.~D.,  2007, \mn@doi [Comput.\ Sci.\ Eng.] {10.1109/MCSE.2007.55}, 9,
  90

\bibitem[\protect\citeauthoryear{{Huterer} \& {Cooray}}{{Huterer} \&
  {Cooray}}{2005}]{2005PhRvD..71b3506H}
{Huterer} D.,  {Cooray} A.,  2005, \mn@doi [\prd] {10.1103/PhysRevD.71.023506},
  \href {http://adsabs.harvard.edu/abs/2005PhRvD..71b3506H} {71, 023506}

\bibitem[\protect\citeauthoryear{{Huterer} \& {Starkman}}{{Huterer} \&
  {Starkman}}{2003}]{2003PhRvL..90c1301H}
{Huterer} D.,  {Starkman} G.,  2003, \mn@doi [\prl]
  {10.1103/PhysRevLett.90.031301}, \href
  {http://adsabs.harvard.edu/abs/2003PhRvL..90c1301H} {90, 031301}

\bibitem[\protect\citeauthoryear{{Jaffe}}{{Jaffe}}{1996}]{1996ApJ...471...24J}
{Jaffe} A.,  1996, \mn@doi [\apj] {10.1086/177950}, \href
  {http://adsabs.harvard.edu/abs/1996ApJ...471...24J} {471, 24}

\bibitem[\protect\citeauthoryear{{Jensen} \& {Nielsen}}{{Jensen} \&
  {Nielsen}}{2007}]{Jensen2001}
{Jensen} F.~V.,  {Nielsen} T.~D.,  2007, {Bayesian Networks and Decision
  Graphs}, 2nd edn.
Springer, New York, \mn@doi{10.1007/978-0-387-68282-2}

\bibitem[\protect\citeauthoryear{{Jordan}}{{Jordan}}{2004}]{graphical}
{Jordan} M.~I.,  2004, \mn@doi [Stat.\ Sci.] {10.1214/088342304000000026}, 19,
  140

\bibitem[\protect\citeauthoryear{{Kelly}, {Hicken}, {Burke}, {Mandel}  \&
  {Kirshner}}{{Kelly} et~al.}{2010}]{2010ApJ...715..743K}
{Kelly} P.~L.,  {Hicken} M.,  {Burke} D.~L.,  {Mandel} K.~S.,   {Kirshner}
  R.~P.,  2010, \mn@doi [\apj] {10.1088/0004-637X/715/2/743}, \href
  {http://adsabs.harvard.edu/abs/2010ApJ...715..743K} {715, 743}

\bibitem[\protect\citeauthoryear{{Kj{\ae}rulff} \& {Madsen}}{{Kj{\ae}rulff} \&
  {Madsen}}{2013}]{km}
{Kj{\ae}rulff} U.~B.,  {Madsen} A.~L.,  2013, {Bayesian Networks and Influence
  Diagrams: A Guide to Construction and Analysis}, 2nd edn.
Springer, New York, \mn@doi{10.1007/978-1-4614-5104-4}

\bibitem[\protect\citeauthoryear{{Kullback} \& {Leibler}}{{Kullback} \&
  {Leibler}}{1951}]{kl}
{Kullback} S.,  {Leibler} R.~A.,  1951, \mn@doi [Ann.\ Math.\ Stat.]
  {10.1214/aoms/1177729694}, 22, 79

\bibitem[\protect\citeauthoryear{{Lago}, {Calv{\~a}o}, {Jor{\'a}s}, {Reis},
  {Waga}  \& {Giostri}}{{Lago} et~al.}{2012}]{2012A&A...541A.110L}
{Lago} B.~L.,  {Calv{\~a}o} M.~O.,  {Jor{\'a}s} S.~E.,  {Reis} R.~R.~R.,
  {Waga} I.,   {Giostri} R.,  2012, \mn@doi [\aap]
  {10.1051/0004-6361/201118599}, \href
  {http://adsabs.harvard.edu/abs/2012A%26A...541A.110L} {541, A110}

\bibitem[\protect\citeauthoryear{{Li}, {Li}, {Wang}  \& {Zhou}}{{Li}
  et~al.}{2016}]{2016MNRAS.460.2586L}
{Li} M.,  {Li} N.,  {Wang} S.,   {Zhou} L.,  2016, \mn@doi [\mnras]
  {10.1093/mnras/stw1063}, \href
  {http://adsabs.harvard.edu/abs/2016MNRAS.460.2586L} {460, 2586}

\bibitem[\protect\citeauthoryear{{Maguire} et~al.,}{{Maguire}
  et~al.}{2012}]{2012MNRAS.426.2359M}
{Maguire} K.,  et~al., 2012, \mn@doi [\mnras]
  {10.1111/j.1365-2966.2012.21909.x}, \href
  {http://adsabs.harvard.edu/abs/2012MNRAS.426.2359M} {426, 2359}

\bibitem[\protect\citeauthoryear{{March}, {Trotta}, {Berkes}, {Starkman}  \&
  {Vaudrevange}}{{March} et~al.}{2011}]{2011MNRAS.418.2308M}
{March} M.~C.,  {Trotta} R.,  {Berkes} P.,  {Starkman} G.~D.,   {Vaudrevange}
  P.~M.,  2011, \mn@doi [\mnras] {10.1111/j.1365-2966.2011.19584.x}, \href
  {http://adsabs.harvard.edu/abs/2011MNRAS.418.2308M} {418, 2308}

\bibitem[\protect\citeauthoryear{{March}, {Karpenka}, {Feroz}  \&
  {Hobson}}{{March} et~al.}{2014}]{2014MNRAS.437.3298M}
{March} M.~C.,  {Karpenka} N.~V.,  {Feroz} F.,   {Hobson} M.~P.,  2014, \mn@doi
  [\mnras] {10.1093/mnras/stt2114}, \href
  {http://adsabs.harvard.edu/abs/2014MNRAS.437.3298M} {437, 3298}

\bibitem[\protect\citeauthoryear{{Mosher} et~al.,}{{Mosher}
  et~al.}{2014}]{2014ApJ...793...16M}
{Mosher} J.,  et~al., 2014, \mn@doi [\apj] {10.1088/0004-637X/793/1/16}, \href
  {http://adsabs.harvard.edu/abs/2014ApJ...793...16M} {793, 16}

\bibitem[\protect\citeauthoryear{{Mukherjee}, {Parkinson}, {Corasaniti},
  {Liddle}  \& {Kunz}}{{Mukherjee} et~al.}{2006}]{2006MNRAS.369.1725M}
{Mukherjee} P.,  {Parkinson} D.,  {Corasaniti} P.-S.,  {Liddle} A.~R.,   {Kunz}
  M.,  2006, \mn@doi [\mnras] {10.1111/j.1365-2966.2006.10427.x}, \href
  {http://adsabs.harvard.edu/abs/2006MNRAS.369.1725M} {369, 1725}

\bibitem[\protect\citeauthoryear{{Nocedal} \& {Wright}}{{Nocedal} \&
  {Wright}}{2006}]{nw}
{Nocedal} J.,  {Wright} S.~J.,  2006, {Numerical Optimization}, 2nd edn.
Springer-Verlag, New York, \mn@doi{10.1007/978-0-387-40065-5}

\bibitem[\protect\citeauthoryear{{Pando}, {Lipa}, {Greiner}  \& {Fang}}{{Pando}
  et~al.}{1998}]{1998ApJ...496....9P}
{Pando} J.,  {Lipa} P.,  {Greiner} M.,   {Fang} L.-Z.,  1998, \mn@doi [\apj]
  {10.1086/305386}, \href {http://adsabs.harvard.edu/abs/1998ApJ...496....9P}
  {496, 9}

\bibitem[\protect\citeauthoryear{{Patil}, {Huard}  \& {Fonnesbeck}}{{Patil}
  et~al.}{2010}]{JSSv035i04}
{Patil} A.,  {Huard} D.,   {Fonnesbeck} C.,  2010, \mn@doi [J.\ Stat.\ Softw.]
  {10.18637/jss.v035.i04}, 35, 1

\bibitem[\protect\citeauthoryear{{P{\'e}rez-Cruz}}{{P{\'e}rez-Cruz}}{2008}]{klknn}
{P{\'e}rez-Cruz} F.,  2008, in {Proc.\ 2008 IEEE International Symposium on
  Information Theory}. Curran Associates, Red Hook, NY, p.~1666,
  \mn@doi{10.1109/ISIT.2008.4595271}

\bibitem[\protect\citeauthoryear{{Perlmutter} et~al.,}{{Perlmutter}
  et~al.}{1999}]{1999ApJ...517..565P}
{Perlmutter} S.,  et~al., 1999, \mn@doi [\apj] {10.1086/307221}, \href
  {http://adsabs.harvard.edu/abs/1999ApJ...517..565P} {517, 565}

\bibitem[\protect\citeauthoryear{{Phillips}}{{Phillips}}{1993}]{1993ApJ...413L.105P}
{Phillips} M.~M.,  1993, \mn@doi [\apjl] {10.1086/186970}, \href
  {http://adsabs.harvard.edu/abs/1993ApJ...413L.105P} {413, L105}

\bibitem[\protect\citeauthoryear{{Phillips}, {Lira}, {Suntzeff}, {Schommer},
  {Hamuy}  \& {Maza}}{{Phillips} et~al.}{1999}]{1999AJ....118.1766P}
{Phillips} M.~M.,  {Lira} P.,  {Suntzeff} N.~B.,  {Schommer} R.~A.,  {Hamuy}
  M.,   {Maza} J.,  1999, \mn@doi [\aj] {10.1086/301032}, \href
  {http://adsabs.harvard.edu/abs/1999AJ....118.1766P} {118, 1766}

\bibitem[\protect\citeauthoryear{{Planck Collaboration}}{{Planck
  Collaboration}}{2015}]{2015arXiv150201589P}
{Planck Collaboration} 2015, preprint, \href
  {http://adsabs.harvard.edu/abs/2015arXiv150201589P} {} (\mn@eprint {arXiv}
  {1502.01589})

\bibitem[\protect\citeauthoryear{{Rao}}{{Rao}}{1945}]{rao}
{Rao} C.~R.,  1945, Sankhy{\=a}, 7, 9

\bibitem[\protect\citeauthoryear{{Riess} et~al.,}{{Riess}
  et~al.}{1998}]{1998AJ....116.1009R}
{Riess} A.~G.,  et~al., 1998, \mn@doi [\aj] {10.1086/300499}, \href
  {http://adsabs.harvard.edu/abs/1998AJ....116.1009R} {116, 1009}

\bibitem[\protect\citeauthoryear{{Riess} et~al.,}{{Riess}
  et~al.}{2007}]{2007ApJ...659...98R}
{Riess} A.~G.,  et~al., 2007, \mn@doi [\apj] {10.1086/510378}, \href
  {http://adsabs.harvard.edu/abs/2007ApJ...659...98R} {659, 98}

\bibitem[\protect\citeauthoryear{{Riess} et~al.,}{{Riess}
  et~al.}{2011}]{2011ApJ...730..119R}
{Riess} A.~G.,  et~al., 2011, \mn@doi [\apj] {10.1088/0004-637X/730/2/119},
  \href {http://adsabs.harvard.edu/abs/2011ApJ...730..119R} {730, 119}

\bibitem[\protect\citeauthoryear{{Rigault} et~al.,}{{Rigault}
  et~al.}{2015}]{2015ApJ...802...20R}
{Rigault} M.,  et~al., 2015, \mn@doi [\apj] {10.1088/0004-637X/802/1/20}, \href
  {http://adsabs.harvard.edu/abs/2015ApJ...802...20R} {802, 20}

\bibitem[\protect\citeauthoryear{{Rubin} et~al.,}{{Rubin}
  et~al.}{2015}]{2015ApJ...813..137R}
{Rubin} D.,  et~al., 2015, \mn@doi [\apj] {10.1088/0004-637X/813/2/137}, \href
  {http://adsabs.harvard.edu/abs/2015ApJ...813..137R} {813, 137}

\bibitem[\protect\citeauthoryear{{Scolnic} et~al.,}{{Scolnic}
  et~al.}{2014}]{2014ApJ...795...45S}
{Scolnic} D.,  et~al., 2014, \mn@doi [\apj] {10.1088/0004-637X/795/1/45}, \href
  {http://adsabs.harvard.edu/abs/2014ApJ...795...45S} {795, 45}

\bibitem[\protect\citeauthoryear{{Shachter}}{{Shachter}}{1998}]{2013arXiv13017412S}
{Shachter} R.~D.,  1998, in {Cooper} G.,  {Moral} S.,  eds, Proc.\ 14th
  Conference on Uncertainty in Artificial Intelligence (UAI-98). Morgan
  Kaufmann, San Francisco, p.~480 (\mn@eprint {arXiv} {1301.7412})

\bibitem[\protect\citeauthoryear{{Shariff}, {Jiao}, {Trotta}  \& {van
  Dyk}}{{Shariff} et~al.}{2016}]{2016ApJ...827....1S}
{Shariff} H.,  {Jiao} X.,  {Trotta} R.,   {van Dyk} D.~A.,  2016, \mn@doi
  [\apj] {10.3847/0004-637X/827/1/1}, \href
  {http://adsabs.harvard.edu/abs/2016ApJ...827....1S} {827, 1}

\bibitem[\protect\citeauthoryear{{Sullivan} et~al.,}{{Sullivan}
  et~al.}{2010}]{2010MNRAS.406..782S}
{Sullivan} M.,  et~al., 2010, \mn@doi [\mnras]
  {10.1111/j.1365-2966.2010.16731.x}, \href
  {http://adsabs.harvard.edu/abs/2010MNRAS.406..782S} {406, 782}

\bibitem[\protect\citeauthoryear{{Suzuki} et~al.,}{{Suzuki}
  et~al.}{2012}]{2012ApJ...746...85S}
{Suzuki} N.,  et~al., 2012, \mn@doi [\apj] {10.1088/0004-637X/746/1/85}, \href
  {http://adsabs.harvard.edu/abs/2012ApJ...746...85S} {746, 85}

\bibitem[\protect\citeauthoryear{{Tegmark}}{{Tegmark}}{1997}]{1997PhRvD..55.5895T}
{Tegmark} M.,  1997, \mn@doi [\prd] {10.1103/PhysRevD.55.5895}, \href
  {http://adsabs.harvard.edu/abs/1997PhRvD..55.5895T} {55, 5895}

\bibitem[\protect\citeauthoryear{{Tegmark}, {Taylor}  \& {Heavens}}{{Tegmark}
  et~al.}{1997}]{1997ApJ...480...22T}
{Tegmark} M.,  {Taylor} A.~N.,   {Heavens} A.~F.,  1997, \mn@doi [\apj]
  {10.1086/303939}, \href {http://adsabs.harvard.edu/abs/1997ApJ...480...22T}
  {480, 22}

\bibitem[\protect\citeauthoryear{{Tonry} et~al.,}{{Tonry}
  et~al.}{2012}]{2012ApJ...750...99T}
{Tonry} J.~L.,  et~al., 2012, \mn@doi [\apj] {10.1088/0004-637X/750/2/99},
  \href {http://adsabs.harvard.edu/abs/2012ApJ...750...99T} {750, 99}

\bibitem[\protect\citeauthoryear{{Tripp}}{{Tripp}}{1998}]{1998A&A...331..815T}
{Tripp} R.,  1998, \aap, \href
  {http://adsabs.harvard.edu/abs/1998A%26A...331..815T} {331, 815}

\bibitem[\protect\citeauthoryear{{Trotta}}{{Trotta}}{2007}]{2007MNRAS.378...72T}
{Trotta} R.,  2007, \mn@doi [\mnras] {10.1111/j.1365-2966.2007.11738.x}, \href
  {http://adsabs.harvard.edu/abs/2007MNRAS.378...72T} {378, 72}

\bibitem[\protect\citeauthoryear{{White} \& {Padmanabhan}}{{White} \&
  {Padmanabhan}}{2015}]{2015JCAP...12..058W}
{White} M.,  {Padmanabhan} N.,  2015, \mn@doi [J.\ Cosmol.\ Astropart.\ Phys.]
  {10.1088/1475-7516/2015/12/058}, \href
  {http://adsabs.harvard.edu/abs/2015JCAP...12..058W} {12, 058}

\bibitem[\protect\citeauthoryear{{Wigner}}{{Wigner}}{1963}]{Wigner:1963:WPM}
{Wigner} E.~P.,  1963, \mn@doi [Can.\ J.\ Math.] {10.4153/CJM-1963-036-4}, 15,
  313

\bibitem[\protect\citeauthoryear{{Wood-Vasey} et~al.,}{{Wood-Vasey}
  et~al.}{2007}]{2007ApJ...666..694W}
{Wood-Vasey} W.~M.,  et~al., 2007, \mn@doi [\apj] {10.1086/518642}, \href
  {http://adsabs.harvard.edu/abs/2007ApJ...666..694W} {666, 694}

\makeatother
\end{thebibliography}

\appendix

\section{Compressed SN Ia data tables}
\label{sec:cdatatab}

We present the compressed JLA $\vvdc{\mu}$ data set in \autoref{tab:compours}
and the covariance matrix in \autoref{tab:covours}
obtained using our method.  They are available at the same 31 redshift locations
as those of \citetalias[table F.1]{2014A&A...568A..22B}.  However, we can also
incorporate the post-compression standardization parameters $(\alpha, \beta,
\Delta_M)$ in our data set.  The mean values of those standardization
parameters are already shown as the first column of \autoref{tab:compcalib},
which can be concatenated with the $\vvdc{\mu}$ vector listed in
\autoref{tab:compours} to form the full compressed data set.  The
standardization parameters are correlated with $\vvdc{\mu}$, a fact reflected
in \autoref{tab:covours}.  It is possible to pre-marginalize over the
standardization parameters before using the compressed data, simply by
dropping the corresponding rows and columns from the tables.

\begin{table*}
    \caption{Compressed JLA SN Ia distance modulus data vector $\vvdc{\mu}$.
	The mean values of post-compression standardization parameters are
	already listed as the first column of \autoref{tab:compcalib}. The
	full table is available online. (Note for arXiv preprint: available as
    ancillary file.)}
    \label{tab:compours}
    \begin{tabular}{cc}
	\hline
	$z$   & $\mu_{\text{dc}}$ \\
	\hline
	0.010 & 33.006 \\
	0.012 & 33.833 \\
	0.014 & 33.862 \\
	0.016 & 34.119 \\
	0.019 & 34.587 \\
	\dots & \dots \\
	1.300 & 44.826 \\
	\hline
    \end{tabular}
\end{table*}

\begin{table*}
    \caption{Joint covariance matrix of compressed JLA distance moduli and
	standardization parameters.  For the purpose of presentation only,
	values in this table have been multiplied by $10^6$, and only the
	upper triangle of the symmetric matrix is shown. The full table
	(without scaling by $10^6$) is available online.  (Note for arXiv
    preprint: available as ancillary file.)}
    \label{tab:covours}
    \begin{tabular}{rrrrrrrr}
	\hline
	 & $\alpha$ & $\beta$ & $\Delta_M$ & $\mu_{\text{dc}, 1}$ &
	$\mu_{\text{dc}, 2}$ & \dots & $\mu_{\text{dc}, 31}$ \\
	\hline
	$\alpha$ & $35$ &  $19$ &  $-30$ &  $11$ &  $-28$ & \dots &  $25$ \\
	$\beta$ & &  $4533$ &  $15$ &  $541$ &  $-577$ & \dots &  $132$ \\
	$\Delta_M$ & & &  $479$ &  $-160$ &  $-117$ & \dots &  $-189$ \\
	$\mu_{\text{dc}, 1}$ & & & &  $20375$ &  $-10398$ & \dots &  $183$ \\
	$\mu_{\text{dc}, 2}$ & & & & &  $27129$ & \dots &  $214$ \\
	\dots & & & & & & \dots & \dots \\
	$\mu_{\text{dc}, 31}$ & & & & & & &  $16300$ \\
	\hline
    \end{tabular}
\end{table*}

\section{Comparing covariance matrices}
\label{apd:covmethod}

Here, we present a method to visually compare covariance matrices of the same
size. Indeed, an element-wise comparison can be done directly.  However it is
possible to design a positive-definite operator that allows for a intuitive
visual comparison.

First let us consider a Gaussian distribution centred around zero with
covariance matrix $\mtx{C}$, $\mathcal{N}(\vv{0}, \mtx{C})$. Let $\mtx{C}_1$
and $\mtx{C}_2$ be covariances for two such distributions.  They are related by
a linear transformation $\mtx{W}_{12}: \vv{x} \sim
\mathcal{N}(\vv{0}, \mtx{C}_1) \to \mtx{W}_{12} \vv{x} \sim \mathcal{N}(\vv{0},
\mtx{C}_2)$, whose matrix representation is the solution to the matrix equation
\begin{equation}
    \mtx{C}_2 = \mtx{W}_{12} \mtx{C}_1 \mtx{W}_{12}^{\intercal}.
    \label{eq:wouter}
\end{equation}

Intuitively, each row of $\mtx{W}_{12}$ can be seen as a window function (dual
vector) being applied on (forming inner product with) random vectors drawn from
the distribution $\mathcal{N}(\vv{0}, \mtx{C}_1)$.  The $k$th window function
maps any of the random vectors to the $k$th component coordinate of the
transformed vectors that follow the target distribution $\mathcal{N}(\vv{0},
\mtx{C}_2)$.  If the target distribution is identical to the original one,
these windows can simply be taken as projections on to coordinate bases, i.e.\
$(1, 0, \dots,0)$, $(0, 1, \dots, 0)$, $\dots (0, 0, \dots 1)$.  Otherwise,
the windows will in general `leak' into other modes unless it is a simple
scaling in one of the dimensions.

However, the matrix $\mtx{W}_{12}$ is not unique.  For example, let the
covariance matrices have factorized form $\mtx{S}_1 \mtx{S}_1^{\intercal} =
\mtx{C}_1$ and $\mtx{S}_2 \mtx{S}_2^{\intercal} = \mtx{C}_2$, where
$\mtx{S}_{1, 2}$ are of the same dimensions as $\mtx{C}_{1, 2}$. Such factors
exist, for example, by the existence of Cholesky factorization or
diagonalizability of positive-definite symmetric matrices.  They are
invertible, for if they are not, we have $\det \mtx{C}_{1, 2} = (\det
\mtx{S}_{1, 2})^2 = 0$.  It follows that any matrix in the form of $\mtx{S}_2
\mtx{P} \mtx{S}_1^{-1}$, where $\mtx{P}$ is a matrix such that $\mtx{P}
\mtx{P}^{\intercal} = \mtx{I}$ (i.e.\ orthogonal), can be a choice for
$\mtx{W}_{12}$.  In the following discussion we aim to find the one that is
positive-definite and symmetric.

Following \citet{1997PhRvD..55.5895T}, \citet{2000MNRAS.312..285H}, and
\citet{2005PhRvD..71b3506H} we can define the matrix square root as
\begin{equation}
    \mtx{C}^{\frac{1}{2}} = \mtx{Q}^{\intercal} \mtx{D}^{\frac{1}{2}} \mtx{Q}
    \label{eq:matrixsqrt}
\end{equation}
for any positive-definite matrix $\mtx{C}$ having eigendecomposition
\begin{equation}
    \mtx{C} = \mtx{Q}^{\intercal} \mtx{D} \mtx{Q}
    \label{eq:eigenfactor}
\end{equation}
where $\mtx{Q}$ is the orthogonal matrix of (row) eigenvectors, $\mtx{D}$ is
the diagonal matrix with positive eigenvalues, and $\mtx{D}^{\frac{1}{2}}$
is the element-wise, positive square root of the matrix diagonal.  It is worth
noting that the eigenvalue decomposition of \eref{eq:eigenfactor} is unique
only up to permutations, but all such permutations map to the same square root
\eref{eq:matrixsqrt}.  Indeed it is the unique positive-definite symmetric
square root of $\mtx{C}$.  The square root defined this way shares eigenvectors
with $\mtx{C}$ and all matrices $\mtx{C}^t = \mtx{Q}^{\intercal} \mtx{D}^t
\mtx{Q}$, where $t \neq 0$ and $\mtx{D}^t$ is the element-wise exponential
function on the diagonal.  All of them are symmetric and positive-definite.

This definition of matrix square root $\mtx{C}^{\frac{1}{2}}$ is related to the
eigenvalue problem
\begin{equation}
    \mtx{C} \vv{x} = \lambda \vv{x} = \lambda \mtx{I} \vv{x}
    \label{eq:eigprob}
\end{equation}
whose solution is \eref{eq:eigenfactor} and the eigenvalues $\lambda$ are
the roots of the characteristic polynomial
\begin{equation}
    p(\lambda) = \det \left( \mtx{C} - \lambda \mtx{I} \right).
\end{equation}
$\mtx{C}^{\frac{1}{2}}$ is the matrix of the mapping $\mtx{W}_{01}$ that takes
$\mathcal{N}(0, \mtx{I})$ to $\mathcal{N}(0, \mtx{C})$, which is a special case
of our problem.  Hence, we would like to find an operator $\mtx{W}_{12}$ that
inherits the properties of this special case.

Motivated by this observation, we can heuristically extend it to the case of a
generalized mapping $\mtx{W}_{12}$ by considering the generalized eigenvalue
problem that extends \eref{eq:eigprob},
\begin{equation}
    \mtx{C}_2 \vv{x} = \lambda' \mtx{C}_1^{-1} \vv{x},
    \label{eq:geneigenprob}
\end{equation}
with the corresponding generalized characteristic polynomial equation
\begin{equation}
    p(\lambda) = \det \left( \mtx{C}_2 - \lambda' \mtx{C}_1^{-1} \right) = 0,
    \label{eq:gencharpoly}
\end{equation}
which has the same roots as
\begin{equation}
    \tilde{p}(\lambda) = \det \left( \mtx{C}_1^{\frac{1}{2}} \mtx{C}_2
    \mtx{C}_1^{\frac{1}{2}} - \lambda' \mtx{I} \right) = 0.
\end{equation}
The matrix $\mtx{C}_1^{\frac{1}{2}} \mtx{C}_2 \mtx{C}_1^{\frac{1}{2}}$, a
product of three positive-definite matrices, is manifestly symmetric, hence
positive-definite \citep[][theorem 2]{Wigner:1963:WPM}. Thus the problem is
reduced to the already solved eigenvalue problem of a positive-definite
symmetric matrix. It follows that there is the eigendecomposition
\begin{equation}
    \mtx{C}_1^{\frac{1}{2}} \mtx{C}_2 \mtx{C}_1^{\frac{1}{2}} =
    \mtx{Q}'^{\intercal} \mtx{D}' \mtx{Q}',
    \label{eq:gendecomp}
\end{equation}
where $\mtx{D}'$ has diagonal elements (generalized eigenvalues) solving
\eref{eq:gencharpoly}.  Again, we use \eref{eq:matrixsqrt} and denote the
`square root' of \eref{eq:gendecomp} as 
\begin{equation}
    \mtx{S}' = \mtx{Q}'^{\intercal} \mtx{D}'^{\frac{1}{2}} \mtx{Q}'.
\end{equation}
Then it follows from \eref{eq:gendecomp} that
\begin{equation}
    \begin{split}
	\mtx{C}_2 = \mtx{C}_1^{-\frac{1}{2}} \mtx{S}' \mtx{S}'
	\mtx{C}_1^{-\frac{1}{2}}
	&= \left( \mtx{C}_1^{-\frac{1}{2}} \mtx{S}' \mtx{C}_1^{-\frac{1}{2}}
	    \right) \mtx{C}_1 \left( \mtx{C}_1^{-\frac{1}{2}} \mtx{S}'
	    \mtx{C}_1^{-\frac{1}{2}} \right) \\
	&= \mtx{W}_{12} \mtx{C}_1 \mtx{W}_{12}^{\intercal},
    \end{split}
\end{equation}
where the manifestly symmetric mapping
\begin{equation}
    \mtx{W}_{12} = \mtx{W}_{12}^{\intercal} = \mtx{C}_1^{-\frac{1}{2}} \left(
    \mtx{C}_1^{\frac{1}{2}} \mtx{C}_2 \mtx{C}_1^{\frac{1}{2}}
    \right)^{\frac{1}{2}} \mtx{C}_1^{-\frac{1}{2}}
    \label{eq:chosenmapping}
\end{equation}
is the matrix we set out to find for \eref{eq:wouter}.

By the aforementioned theorem of \citet{Wigner:1963:WPM}, $\mtx{W}_{12}$ itself
is positive-definite.  Notice that in the expression \eref{eq:chosenmapping}
the matrix exponent $1/2$ cannot simply be distributed to the individual
factors of the matrix product $\mtx{C}_1^{\frac{1}{2}} \mtx{C}_2
\mtx{C}_1^{\frac{1}{2}}$.  Instead, it must be understood by solving the
generalized eigenvalue problem of \eref{eq:geneigenprob}.

Just like the square root defined in \eref{eq:matrixsqrt} tends to conserve
the window function bandwidth \citep{1997PhRvD..55.5895T}, the extension
$\mtx{W}_{12}$ as defined in \eref{eq:chosenmapping} also creates narrow
windows. In other words, it is not likely to generate a combination of wide
windows in order to account for a small difference.  This is a desirable
feature for the matrices we want to compare, because we expect small
differences, some of which are simply numerical artefacts of the computation.

\bsp	
\label{lastpage}
\end{document}